\documentclass[10pt,english,pre,nofootinbib,reprint,showpacs,showkeys,preprintnumbers,floatfix]{revtex4-1}
\usepackage[latin9]{inputenc}
\usepackage[a4paper]{geometry}
\usepackage{babel}
\usepackage{units}
\usepackage{amsmath}
\usepackage{amssymb}
\usepackage{graphicx}
\usepackage{esint}
\usepackage[unicode=true,
 bookmarks=true,bookmarksnumbered=false,bookmarksopen=false,
 breaklinks=false,pdfborder={0 0 1},backref=false,colorlinks=false]
 {hyperref}
\hypersetup{pdftitle={Improving self calibration},
 pdfauthor={Torsten En{\ss}lin et al.}}

\makeatletter

\providecommand{\tabularnewline}{\\}

\newcommand{\lyxdeleted}[3]{}

 \@ifundefined{textcolor}{}
 {%
   \definecolor{BLACK}{gray}{0}
   \definecolor{WHITE}{gray}{1}
   \definecolor{RED}{rgb}{1,0,0}
   \definecolor{GREEN}{rgb}{0,1,0}
   \definecolor{BLUE}{rgb}{0,0,1}
   \definecolor{CYAN}{cmyk}{1,0,0,0}
   \definecolor{MAGENTA}{cmyk}{0,1,0,0}
   \definecolor{YELLOW}{cmyk}{0,0,1,0}
 }

\usepackage{aas_macros}
\usepackage{bbold}
\newcommand{\R}{B}

\renewcommand{\lyxdeleted}[3]{}

\makeatother

\begin{document}

\title{Improving self-calibration}

\author{Torsten A. Enßlin, Henrik Junklewitz, Lars Winderling, Maksim Greiner,
Marco Selig}

\affiliation{{\small Max-Planck-Institut für Astrophysik, Karl-Schwarzschildstr.
1, 85748 Garching, Germany}}

\affiliation{Ludwig-Maximilians-Universität München, Geschwister-Scholl-Platz
1, 80539 Munich, Germany}
\begin{abstract}
Response calibration is the process of inferring how much the measured
data depend on the signal one is interested in. It is essential for
any quantitative signal estimation on the basis of the data. Here,
we investigate self-calibration methods for linear signal measurements
and linear dependence of the response on the calibration parameters.
The common practice is to augment an external calibration solution
using a known reference signal with an internal calibration on the
unknown measurement signal itself. Contemporary self-calibration schemes
try to find a self-consistent solution for signal and calibration
by exploiting redundancies in the measurements. This can be understood
in terms of maximizing the joint probability of signal and calibration.
However, the full uncertainty structure of this joint probability
around its maximum is thereby not taken into account by these schemes.
Therefore better schemes -- in sense of minimal square error -- can
be designed by accounting for asymmetries in the uncertainty of signal
and calibration. We argue that at least a systematic correction of
the common self-calibration scheme should be applied in many measurement
situations in order to properly treat uncertainties of the signal
on which one calibrates. Otherwise the calibration solutions suffer
from a systematic bias,\lyxdeleted{Torsten Enlin,,,}{Sat Sep  6 16:58:18 2014}{
} which consequently distorts the signal reconstruction. Furthermore,
we argue that non-parametric, signal-to-noise filtered calibration
should provide more accurate reconstructions than the common bin averages
and provide a new, improved self-calibration scheme. We illustrate
our findings with a simplistic numerical example.
\end{abstract}

\pacs{89.70.Eg, 11.10.-z, 02.50.Tt, 06.20.fb, 07.05.Kf}

\keywords{Information theory -- Field theory -- Inference methods -- Standards
and calibration -- Data analysis: algorithms and implementation}

\maketitle

\section{Introduction}

\subsection{Motivation }

Any measurement device needs a proper calibration, otherwise an accurate
translation of the raw measurement data into a common system of units
is impossible. Our ability to process, combine, communicate, and draw
conclusions from the results of measurements depends\lyxdeleted{Torsten Enlin,,,}{Sat Sep  6 16:58:18 2014}{
} critically on the achieved calibration accuracy. 

The calibration problem is widespread across different fields. Knowing
the amplifier gain factors and detector efficiencies of physical measurement
apparatuses is necessary to analyze their data. In astronomy, the
point spread function of a telescope observation might be unknown,
since it could depend on varying atmospheric influences. In analyzing
sociological questionnaires, the reliability of people's answers might
differ from topic to topic, but needs to be taken into account. In
all those cases, the measurement response to the quantity of interest,
our signal, needs to be known. This response expresses how the data
reacts (on average) to changes in the\lyxdeleted{Torsten Enlin,,,}{Sat Sep  6 16:58:18 2014}{
} signal. Only if one knows the response precisely, one can accurately
recover the signal of interest correctly from data. The process of
the response determination is called \emph{calibration}, its result
the \emph{calibration solution}, \emph{calibration reconstruction},
or just \emph{calibration} for brevity.

Several kinds of calibration uncertainties appear in practice: offsets
(or additive noise), gain uncertainties (or multiplicative noise),
and nonlinearities (e.g. receiver saturation). This work deals with
the first two kinds of problems, multiplicative and additive noise.
Noise denotes here any influence of the data which is not due to the
signal of interest, be it stochastic or just unknown in nature. Non-linear
signal responses complicate the signal inference considerably. If
the non-linearities are known, the generic insights about the calibration
of additive and multiplicative noise derived in this work still apply.
The calibration of unknown non-linearities is beyond the scope of
this paper, though.

The classical way to calibrate a measurement device is to apply it
to a known reference signal, the calibrator. The obtained instrument
response to this can then be used to gauge the instrument and to interpret
the data obtained from measuring an unknown signal \citep{scheffe1973statistical,1982JRSSB44...287,1991ISR59...309O,zbMATH01006589,feudale2002transfer,2008ApJ...674.1217P}. 

However, in many measurement situations the response depends strongly
on time, location, temperature, energy, frequency, or other dimensions.
A simultaneous measurement of both calibrator and signal is often
impossible. The external calibration needs then to be extrapolated
within the time (space, energy, ...) domain of the signal measurement.
Extrapolation in time is only possible if the calibration exhibits
sufficient auto-correlation. This auto-correlation could be used to
optimally filter out noise in the calibration solution. In practice,
however, usually only averaging of the individual calibration solutions
within suitably chosen intervals is performed. 

An calibration obtained might be further improved by exploiting redundancies
in the signal measurement. If the same aspects of the signal are measured
repeatedly, but the data show significant deviations between the individual
measurements, this indicates a change in the instrument's sensitivity.
Thus, an external calibration can often be improved by an internal
calibration using the unknown signal itself as an additional source
of calibration information.

The usual internal calibration or self-calibration (\emph{selfcal}
\citep{1967JSpRo...4..554J,1980SPIE..231...18S,1981MNRAS.196.1067C,1982syma.work...13C,1984ARA&A..22...97P,1984sdra.conf..154E,2010A&A...524A..61N,2011AN....332..759L})
scheme proceeds as follows. A coarse external calibration is obtained
and then applied to the data to get a first signal reconstruction.
This signal reconstruction is then used for a refinement of the calibration,
which in turn helps to further improve the signal reconstruction.
The reconstruction and calibration operations are repeated until some
desired convergence criteria are met. 

It is, in general, unclear whether such a procedure converges and
whether the obtained solution is reasonable. There are \emph{selfcal}
schemes derived from minimizing an objective function \citep{2008ApJ...674.1217P}
and convergence can be proven for them. However, for empirically designed
\emph{selfcal} schemes, as used, e.g., in radio interferometry, such
a proof is often missing. It could well be that only a self-consistent
solution of an incorrect signal and an incorrect calibration is obtained,
although the joint fit to the data is perfect.

This problem is of generic nature. If a measured datum depends on
two unknowns, the signal and the instrument sensitivity, these cannot
unambiguously be reconstructed. The additional presence of measurement
noise makes this inference problem even harder. External calibration
is essential, but often relies on the ability to extrapolate it into
domains in time or location, where -- strictly speaking -- it was
not measured for. 

In this work, we show that the classical \emph{selfcal} scheme can
be understood as a joint maximization of the joint posterior probability
of signal and calibration given the data. This posterior represents
all available information on signal and calibration. A stable fix
point of the \emph{selfcal} scheme is a maximum of this joint posterior.
It therefore represents the most likely combination of signal and
calibration, at least in some vicinity of the fix point.

However, such Maximum A Posteriori (MAP) estimators are known to be
prone to over-fitting the data. A posterior mean signal would be optimal
with respect to an expected square error norm \citep[e.g.][]{2009PhRvD..80j5005E}.
In case of a symmetric posterior, mean and maximum coincide and the
MAP estimator is also optimal in this sense. However, the presence
of a nuisance parameter, here being the unknown calibration, can turn
an originally symmetric problem into a skewed one. As a consequence,
the maximum of such a skewed, non-symmetric posterior is systematically
biased away from the location of the posterior mean \citep[e.g.][]{2011PhRvD..83j5014E}.
Indeed, we will show in this work that using the joint MAP estimator
of signal and calibration, as the \emph{selfcal} scheme does, implies
a systematical bias with respect to the more optimal posterior mean
of signal and calibration.

\subsection{Previous work}

The previous work on calibration is vast, in particular the mathematical-statistical
literature. It may be classified into whether it deals with univariate
or multivariate calibration problems, concentrates on external or
internal calibration, and uses frequentist or Bayesian methodologies.
A review of various mathematical treatments of external calibration
(uni- and multivariant as well as frequentistic and Bayesian) can
be found in \citep{1991ISR59...309O}. 

External calibration means that an external, high-quality dataset
is used to map out and reconstruct the response of a measurement device.
This could be a single real function (univariate calibration, e.g.
\citep{scheffe1973statistical}) or a vector valued function (multivariate
calibration, e.g. \citep{1982JRSSB44...287}). This calibrated response
is then used in the interpretation of the following measurements.
The main challenge in external calibration is to construct a noise
suppressing reconstruction operation, which takes the often unknown
noise variance properly into account. The function might be of parametric
form \citep{scheffe1973statistical}, or non-parametric estimators
might be used \citep{zbMATH01006589,2012PhRvD..86a5027D}. It was
realized that in many situations a calibration obtained at some instant
is not accurate for subsequent measurements, as the instrument might
have changed with time. An appropriate calibration transfer method
should be employed that takes such uncertainties properly into account
\citep{feudale2002transfer}. 

Internal calibration deals with the situation that an external calibration
solution is not available, or is known to be inaccurate. For example,
the instrument response might change on timescales comparable to the
one needed to switch the instrument to the calibrator signal. This,
for example, is a common problem in radio interferometry, where the
rapidly changing Earth ionosphere can be regarded as part of the telescope
optics. In such cases, the signal of interest has to serve as a calibration
signal as well. The resulting \emph{selfcal} schemes image the signal
with an assumed calibration, calibrate on this signal reconstruction,
and repeat these operations until convergence \citep{1967JSpRo...4..554J,1980SPIE..231...18S,1981MNRAS.196.1067C,1982syma.work...13C,1984ARA&A..22...97P,1984sdra.conf..154E,2010A&A...524A..61N,2011AN....332..759L}.
To the knowledge of the authors, an information theoretical investigation
is lacking about under which conditions this leads to reliable results,
and when it fails, although practitioners certainly have developed
a good intuition on this. 

A rigorous information theoretical treatment of the problem of unknown
calibration should be build on the calibration marginalized likelihood,
since it contains all the available information from the data and
on the measurement process. For a measurement with Gaussian noise,
linear response, and linear calibration uncertainties with a Gaussian
distribution of known covariance this marginal likelihood can be calculated
analytically \citep{2002MNRAS.335.1193B} and is reproduced here in
Eq.~\eqref{eq:marginal likelihood}. This likelihood is a Gaussian
probability density in the data, with a signal dependent covariance.
Thus, the resulting signal posterior is very non-Gaussian. If the
mean of this signal posterior can be calculated, all the available
internal calibration information is taken implicitly into account,
and there is no need for a determination of the calibration. In case
of non-parametric measurement and calibration problems, the dimensionality
of the problem is, however, often too large (virtually unbound) for
the usual Monte-Carlo methods to sample the posterior. To tackle such
and other problems information field theory \citep{1999physics..12005L,2009PhRvD..80j5005E,2013AIPC.1553..184E}
was developed. This exploits the mathematical and conceptual similarities
of the non-parametric inference problem with statistical field theories
well-known in mathematical physics. For example, the reconstruction
of Gaussian random fields with unknown covariance, as also needed
for calibration, was successfully treated in this framework \citep{2011PhRvD..83j5014E,2011PhRvE..84d1118O,2012A&A...542A..93O-short,2012PhRvD..86a5027D,2013PhRvE..87c2136O}.

For the effective treatment of non-Gaussian posteriors the method
of minimal Gibbs free energy \citep{2010PhRvE..82e1112E} (a thermodynamical
incarnation of the variational Bayes approach) has proven to be useful
and was applied to the calibration marginalized signal inference problem
by Ref. \citep{2012MasterWinderling}. However, due to the contrived
structure of the marginal likelihood, relatively coarse approximations
had to be used there to get to analytical formulas. For this reason,
a more pragmatic approach shall be followed here in tackling the internal
calibration problem.

\subsection{Structure of this work}

This work is organized as follows. In order to develop an intuitive
understanding, we investigate an illustrative example from a frequentist
and Bayesian perspective in Sec.~\ref{sec:Illustrative-example}.
Then, we investigate in Sec.~\ref{sec:Theory-of-calibration}\lyxdeleted{Torsten Enlin,,,}{Sat Sep  6 16:58:18 2014}{
} the general theory of calibration of linear measurements with partly
unknown response operators, in particular external calibration, classical
\emph{selfcal}, and a new, uncertainty corrected \emph{selfcal} schemes.
These different approaches are compared in Sec.~\ref{sec:Numerical-example}
via a numerical example that is based on the illustrative example
of Sec.~\ref{sec:Illustrative-example}. We conclude in Sec.~\ref{sec:Conclusions}
with a summary of our main findings and a brief perspective on what
would be required to develop a full theory of calibration.

\section{Illustrative example\label{sec:Illustrative-example}}

It is the goal of this work to improve the present \emph{selfcal}
schemes such that the reconstructed signal is closer to the a posteriori
mean. It turns out that this is a problem of high mathematical complexity
even for linear responses. The most important correction we find can
be understood intuitively, though. For this, we first turn to a simplistic
example, which we investigate using a less formal language. A more
general and rigorous treatment will be given in Sec.~\ref{sec:Theory-of-calibration},
which is able to deal with the complex linear responses one can find
in practice, like convolving telescope beams etc. The illustrative
example introduced here will be simulated in Sec. \ref{sec:Numerical-example}
and is also the basis of the figures in this article.

A signal $s$ should be $ $observed with an instrument that has a
sensitivity or gain $g$. In our illustrative example, which will
be replaced by a more general case later on, the instrument's data,
\begin{equation}
d=g\, s+n,\label{eq:d=00003Dgs+n}
\end{equation}
is further corrupted by noise $n$. Here and in the following, any
calibration offset in the data is regarded as part of the noise $n$.

Signal, noise, and gains should be independent stochastic processes
so that their joint probability separates according to 
\begin{equation}
\mathcal{P}(n,\, g,\, s)=\mathcal{P}(n)\,\mathcal{P}(g)\,\mathcal{P}(s).\label{eq:independence}
\end{equation}

At this stage, the problem of signal and gain reconstruction is symmetrically
degenerate, since we know as much about the signal as about the gain
given the data. Typically, the gain is not completely unknown, for
example, the sign of the instrument gain is usually known. For definiteness,
let us assume it to be positive and actually $g=1+\gamma$, with $1$
being the known part of the gain and $\gamma$ denoting the unknown
part that we need to calibrate. This could as well be positive as
negative with the same probability. We will refer to any estimate
of $\gamma$ as a \emph{calibration}.

\subsection{Frequentist perspective}

In frequentist data analysis, repeated instances of the data are assumed
to exist. These permit to perform data averages that can be tailored
towards statistical averages. We adopt for a moment this perspective,
since it allows us to highlight the essence of the calibration problem. 

If we would know the calibration, we could infer the signal by averaging
over the data in a way that averages over noise realizations, 
\begin{equation}
\langle d\rangle_{(n|\gamma,s)}=(1+\gamma)\, s+\underbrace{\langle n\rangle_{(n)}}_{=0}.\label{eq:d_avn}
\end{equation}
Here, we assumed the noise to have a zero mean and denote averages
over the probability of $a$ given $b$ by 
\begin{equation}
\langle f(a)\rangle_{(a|b)}\equiv\int\mathcal{D}a\, f(a)\, P(a|b).\label{eq:average}
\end{equation}
Here $\int\mathcal{D}a$ denotes the phase space integral of $a$,
at the moment a finite dimensional integral like $\int dn,$ and later
on also path-integrals over functional spaces.

In case we do not know the calibration, we could still learn something
about the signal, if we are able to average over the data in a way
that averages over noise and calibration realizations. This reveals
the signal, since
\begin{equation}
\langle d\rangle_{(n,\gamma|s)}=(1+\underbrace{\langle\gamma\rangle_{(\gamma)}}_{=0})\, s+\underbrace{\langle n\rangle_{(n)}}_{=0}=s.
\end{equation}

This is less sensitive to the signal since we need more data to perform
our averaging of two stochastic processes, noise and calibration.
But the point we want to make is that the signal can be estimated
from a suitable linear data average, even without knowing the precise
calibration, since there is a known and positive part of the response
of the data to the signal. 

Obtaining information on the unknown calibration $\gamma$, which
would help us to get the signal more accurately, is more difficult.
If we want to perform an analogous averaging to retrieve some information
on $\gamma$, now over noise and signal realizations, we find 
\begin{equation}
\langle d\rangle_{(n,s|\gamma)}=(1+\gamma)\,\underbrace{\langle s\rangle_{(s)}}_{=0}+\underbrace{\langle n\rangle_{(n)}}_{=0}=0,
\end{equation}
while assuming a zero mean for the signal as well. Thus, at linear
order in the data, there is no calibration information available.
We cannot proceed without some knowledge of the signal since the response
of the data to our calibration could as well be positive (for $s>0$)
as negative (for $s<0$). Furthermore, whenever the signal is close
to zero, the data respond only poorly to the calibration. 

In \emph{selfcal} we obtain some information on the signal from the
data, e.g., by using only the known part of the response, which then
might be used to analyze the data for a better guess on the calibration.
This means the data have to be used at least twice (first a rough
signal reconstruction, then calibrating on this) and we end up with
a scheme that is at least quadratically in the data. Indeed, if we
investigate averages over squared data,
\begin{equation}
\langle d^{2}\rangle_{(n,s|\gamma)}=\langle s^{2}\rangle_{(s)}(1+\gamma)^{2}+\langle n^{2}\rangle_{(n)},\label{eq:d^2}
\end{equation}
we find that this contains terms that are directly sensitive to the
calibration and therefore calibration information is available. 

It should be noted that the sensitivity of this squared data to the
gains depends on the signal (and noise) variance, which we therefore
would need to know. Any systematic error in its determination from
data will lead to a systematic bias in the calibration. If such a
biased calibration is used again for improving the signal variance
in an attempt to iteratively improve the calibration solution, the
bias is even increased. Without any external calibration constraints,
the \emph{selfcal} solution would easily drift far away from an initially
acceptable calibration.%
\footnote{This instability is well known in radioastronomical interferometry.
To suppress it, it is common practice to apply \emph{selfcal} only
to either the phases of the complex gain coefficients and to keep
the gain amplitudes fixed or vice versa. %
} Thus, a strong, but self-consistent bias can be present in the results
of \emph{selfcal}.

In practice, \emph{selfcal} is rarely done using Eq.~\eqref{eq:d^2}
as this requires too much data with comparable calibration coefficients
for getting reliable averages to measure the calibration from the
data variance. More direct and more sensitive calibration methods
are used, e.g.\ the mentioned iteration of reconstruction and calibration
steps. The \emph{selfcal} instability exists there as well, in a slightly
more subtle form. For the detailed information theoretical development
and investigation of such methods, we switch now to a Bayesian perspective.

\subsection{Bayesian perspective}

In probabilistic logic (see, e.g., Refs.~\citep{Cox1963,2003prth.book.....J,2008arXiv0808.0012C}),
only a single realization of a data set needs to be available. All
reasoning has to be done conditional to these data and averages over
different data realizations are not part of the resulting data analysis
method. Probabilities express the strength of believe in a certain
possibility conditionally that some other statement is assumed to
be true and not necessarily how often this possibility happens to
be the case as in frequentist thinking. The data are regarded as a
vector of values $d=(d_{1},\ldots,\, d_{\mathtt{n}})\in\mathbb{R}^{\mathtt{n}}$,
$\mathtt{n}\in\mathbb{N}$, for which any datum $d_{i}$ could be
the result of an unique, non-reproducible measurement, as e.g.~its
gain $g_{i}=1+\gamma_{i}$ probably never takes exactly the same value
again.%
\footnote{Repeated measurements or measurements with different instruments can
be combined into a single data vector by simple concatenation of the
individual data vectors.%
} 

The measurement equation of our illustrative example is still Eq.~\eqref{eq:d=00003Dgs+n}
if we read it as a vector equation with components
\begin{equation}
d_{i}=(1+\gamma_{i})\, s_{i}+n_{i}.\label{eq:di}
\end{equation}

We might want to calculate the signal averaged over all unknowns,
but conditioned to the data,
\begin{eqnarray}
m & = & \langle s\rangle_{(n,\gamma,s|d)},\label{eq:m Bayesian}
\end{eqnarray}
since this is known (see, e.g., Ref.~\citep{2009PhRvD..80j5005E})
to minimize the expected square error 
\begin{equation}
\langle(s-m)^{2}\rangle_{(n,\gamma,s|d)}.\label{eq:square error}
\end{equation}

On linear order in the data, the optimal estimator of this mean is
known to be given by (see e.g. \citep{1995ApJ...449..446Z})
\begin{equation}
m=\langle s\, d^{\dagger}\rangle_{(n,\gamma,s)}\langle d\, d^{\dagger}\rangle_{(n,\gamma,s)}^{-1}d+\mathcal{O}(d^{2})
\end{equation}

with
\begin{eqnarray}
\langle s\, d^{\dagger}\rangle_{(n,\gamma,s)} & = & \underbrace{\langle s\, s^{\dagger}\rangle_{(s)}}_{\equiv S}\mbox{ and}\\
\langle d_{i}\overline{d_{j}}\rangle_{(n,\gamma,s)} & = & (1+\underbrace{\langle\gamma_{i}\overline{\gamma_{j}}\rangle_{(\gamma)}}_{\equiv\Gamma_{ij}})\, S_{ij}+\underbrace{\langle n_{i}\overline{n_{j}}\rangle_{(n)}}_{\equiv N_{ij}},\nonumber 
\end{eqnarray}
where the bar denotes complex conjugation. Here we defined the matrices
$S=\langle s\, s^{\dagger}\rangle_{(s)}$, $\Gamma=\langle\gamma\,\gamma^{\dagger}\rangle_{(\gamma)}$,
and $N=\langle n\, n^{\dagger}\rangle_{(n)}$ that express the a priori
uncertainty covariances in signal, calibration and noise, as well
as the notation $\dagger$ for the transpose of a vector (and its
complex conjugate in case it is a complex number). This optimal linear
estimator is known under many names, like Minimal Square Error (MSE)
estimator, generalized Wiener filter \citep{1949wiener}, and others.

Using matrix notation and defining the component-wise matrix product
$(S*\Gamma)_{ij}=S_{ij}\Gamma_{ij}$ (no summation), we get
\begin{equation}
m\approx\underbrace{S\,\left[S+S*\Gamma+N\right]^{-1}}_{F}d\label{eq:linear data filter}
\end{equation}
and find that the reconstruction is a filtered version of the data.
The filter $F$ reduces the variance since its ``denominator'' $ $$S+S*\Gamma+N$
is spectrally%
\footnote{Meaning that $\xi^{\dagger}(S+S*\Gamma+N)\,\xi>\xi^{\dagger}S\,\xi$
for $\forall\xi\in\mathbb{R}^{\mathtt{n}}\backslash$0.%
} larger than the ``numerator'' $S$. We can write $F<\mathbb{1}$
(spectrally). As larger the noise variance $N$ is with respect to
the signal variance $S$, as stronger the down-weighting of the data.
Further down-weighting comes from the combined signal and calibration
variation $S*\Gamma$. However, if $N\ll S$ and $\Gamma\ll\mathbb{1}$
(spectrally) the filter is close to the identity $\mathbb{1}$ and
the signal estimate $F\, d$ is nearly unfiltered data. 

In any case, the expected covariance of this reconstruction,
\begin{equation}
\langle mm^{\dagger}\rangle_{(n,s,\gamma)}=S\,\left[S+S*\Gamma+N\right]^{-1}S=F\, S<S,\label{eq:m variance}
\end{equation}
is (spectrally) smaller than that of the signal, S. 

Using this linear data filter $F$, Eq.~\eqref{eq:linear data filter},
is in general not a bad idea since it is a conservative approach to
signal reconstruction under calibration uncertainties. It adds the
impact of the calibration uncertainty $S*\Gamma$ to the noise budget
$N$ of a generalized Wiener filter \citep{1949wiener}, which is
then applied to the data. The disadvantage of this approach is that
even in case the signal is so strong that the signal-to-noise ratio
is excellent, $S\gg N$ (spectrally), the calibration-noise covariance,
$S*\Gamma$, can still be substantial as it increases with increasing
signal strength. For large calibration uncertainties, better, and
necessarily non-linear methods have to be used, since among all possible
linear data filters, $F$ is already the optimal one (in the sense
of minimizing Eq.~\eqref{eq:square error}). 

The optimal (non-linear) method can be constructed by rewriting \eqref{eq:m Bayesian}
as
\begin{eqnarray}
m & = & \int\!\mathcal{D}n\int\!\mathcal{D}\gamma\int\!\mathcal{D}s\,\mathcal{P}(n,\gamma,s|d)\, s\nonumber \\
 & = & \int\!\mathcal{D}\gamma\int\!\mathcal{D}s\,\mathcal{P}(\gamma,s|d)\, s\nonumber \\
 & = & \int\!\mathcal{D}\gamma\,\mathcal{P}(\gamma|d)\int\!\mathcal{D}s\,\mathcal{P}(s|d,\,\gamma)\, s\nonumber \\
 & = & \langle\langle s\rangle_{(s|d,\gamma)}\rangle_{(\gamma|d)}.\label{eq:m Bayesian-1}
\end{eqnarray}
Here, we have performed a noise marginalization%
\footnote{This marginalization is trivial since the term $\mathcal{P}(d|n,\gamma,s)=\delta(d-(1+\gamma)\, s-n)$
can be obtained using Bayes theorem, which cancels the noise phase
space integral $\int\mathcal{D}n$.%
} and have split $ $ $\mathcal{P}(s,\gamma|d)$ via the product rule
into $\mathcal{P}(s|d,\gamma)\,\mathcal{P}(\gamma|d)$. The inner
signal average $\langle s\rangle_{(s|d,\gamma)}$ assumes the calibration
to be known. However, the outer average goes over the unknown calibration
while weighting each possible calibration according to its posterior
probability given the data, $\mathcal{P}(\gamma|d)$.

The inner signal average might in many situations be well dealt with
by using the optimal linear estimator, 
\begin{eqnarray}
\langle s\rangle_{(s|d,\gamma)} & \approx & \langle s\, d^{\dagger}\rangle_{(n,s|\gamma)}\langle d\, d^{\dagger}\rangle_{(n,s|\gamma)}^{-1}d\nonumber \\
 & = & \underbrace{S\,\mathrm{R^{\dagger}}\left[R\, S\, R^{\dagger}+N\right]^{-1}}_{\equiv W}d,\label{eq:Wiener filter data space}
\end{eqnarray}
where $R=\mathrm{diag}(1+\gamma)$ is the calibration dependent response
matrix of our measurement. The previously problematic signal suppression
by the term $S*\Gamma$ in Eq.~\eqref{eq:linear data filter} became
more specific, since $R\, S\, R^{\dagger}=S+S*(\gamma\,\gamma^{\dagger})$
and therefore $\Gamma\rightarrow\gamma\,\gamma^{\dagger}$. We therefore
expect to obtain a higher fidelity signal recovery even when the subsequently
applied calibration averaging in Eq.~\eqref{eq:m Bayesian-1} might
smooth out some of the features present in $\langle s\rangle_{(s|d,\gamma)}$
as the posterior average $\langle\gamma\gamma^{\dagger}\rangle_{(\gamma|d)}$
implies much less averaging than the prior average $\Gamma=\langle\gamma\gamma^{\dagger}\rangle_{(\gamma)}$.

It is common practice to use a single ``best'' calibration solution
$\gamma^{\star}$, a so-called \textit{point estimate}, instead of
averaging over all possible calibrations. Thus, implicitly $\mathcal{P}(\gamma|d)\approx\delta(\gamma-\gamma^{\star})$
is assumed. This is indeed often a good approximation, as we will
argue in Sec.~\ref{sub:Calibration-marginalized-imaging}. The next
order corrections that take into account the width of the distribution
$\mathcal{P}(\gamma|d)$ are usually small. An imperfectly chosen
$\gamma^{\star}$ has typically a larger impact on the reconstruction
quality than these corrections and our focus should, therefore, be
on how to calibrate most reliably. 

Again, we regard the posterior mean as a good estimate choose it as
a starting point, 
\begin{equation}
\gamma^{\star}=\langle\gamma\rangle_{(s,\gamma|d)}=\langle\langle\gamma\rangle_{(\gamma|d,s)}\rangle_{(s|d)}.\label{eq:correct gamma average}
\end{equation}

If we would know the signal close enough, calibration would be simple,
as we could ignore the outer averaging over $\mathcal{P}(s|d)$. We
would form signal subtracted data $d'=d-s=\gamma\, s+n$ from which
we could construct the optimal linear estimator of the calibration,
\begin{eqnarray}
\gamma^{\star} & \approx & \langle\gamma\rangle_{(\gamma|d',s)}\approx\langle\gamma\, d'^{\dagger}\rangle_{(n,\gamma|s)}\langle d'd'^{\dagger}\rangle_{(n,\gamma|s)}^{-1}d'\nonumber \\
 & = & \Gamma\,\mathrm{R'^{\dagger}}\left[R'\,\Gamma\, R'^{\dagger}+N\right]^{-1}(d-s),\label{eq:Wiener filter for gamma}
\end{eqnarray}
 with $R'=\mathrm{diag}(s)$ being the response of the data to the
calibration parameters $\gamma$.

Iterating the linear estimators for signal and calibration, \eqref{eq:Wiener filter data space}
and \eqref{eq:Wiener filter for gamma} while assuming $s=W\, d$
and $\gamma=\gamma^{\star}$, is then a plausible \emph{selfcal} scheme.
It ignores, however, the uncertainties in signal reconstruction and
calibration and therefore might suffer from a bias similar to the
one discussed before using frequentist arguments. In particular, the
outer averaging in Eq.~\eqref{eq:correct gamma average} over $\mathcal{P}(s|d)$
is crucial. If we ignore for a moment, for simplicity, the signal
dependence of the ``denominator'' in Eq.~\eqref{eq:Wiener filter for gamma}
we see that our calibration estimator
\begin{eqnarray}
\gamma^{\star} & = & \langle\Gamma\,\underline{\mathrm{R'^{\dagger}}}\left[\ldots\right]^{-1}(d-\underline{s})\rangle_{(s|d)}\nonumber \\
 & \approx & \mathcal{O}(\langle s\rangle_{(s|d)})-\mathcal{O}(\langle s\, s^{\dagger}\rangle_{(s|d)})
\end{eqnarray}
requires the knowledge of the a posteriori signal mean $m=\langle s\rangle_{(s|d)}$
and variance $\langle s\, s^{\dagger}\rangle_{(s|d)}$ since the two
underlined ``numerator'' terms both contain the unknown signal.
In classical \emph{selfcal} schemes $\langle s\, s^{\dagger}\rangle_{(s|d)}$
is approximated by $m\, m^{\dagger}$. The latter has, however, less
variance than the former if $m$ is a filtered version of $s$ as
assumed here%
\footnote{See Eq.~\eqref{eq:m variance}, which is valid also here if we set
$S*\Gamma\rightarrow0$ there. If $m$ resulted from naive data averaging,
noise remnants might be significantly present and can lead to an overestimation
of the posterior signal variance and therefore also to a systematically
biased calibration. %
}. This means that a systematic bias is present in such schemes, as
the $\mathcal{O}(\langle s\, s^{\dagger}\rangle_{(s|d)})$ term is
systematically underestimated by $m\, m^{\dagger}$ leading to an
overestimation of $\gamma^{\star}$. The most important result of
this work is to show how to correct for this bias. 

Such a bias was not present in case of the signal estimation using
a point estimator $\gamma^{\star}$ for the calibration (instead of
the calibration averaging). The difference lies in the fact that for
the chosen illustrative data model, Eq.~\eqref{eq:di} (as well as
for many realistic measurement situations), the symmetry of signal
and gain as suggested by Eq.~\eqref{eq:d=00003Dgs+n} is broken,
since $s$ varies around zero and $g$ around a known non-zero value.
Thus, signal estimators can be built on a more reliably non-zero gain
than calibration estimators, which have to exploit opportunistically
any sufficiently non-zero signal fluctuation suitable for calibration.

\section{Theory of calibration\label{sec:Theory-of-calibration}}

\subsection{Generic problem}

\begin{figure*}
\includegraphics[width=1\textwidth]{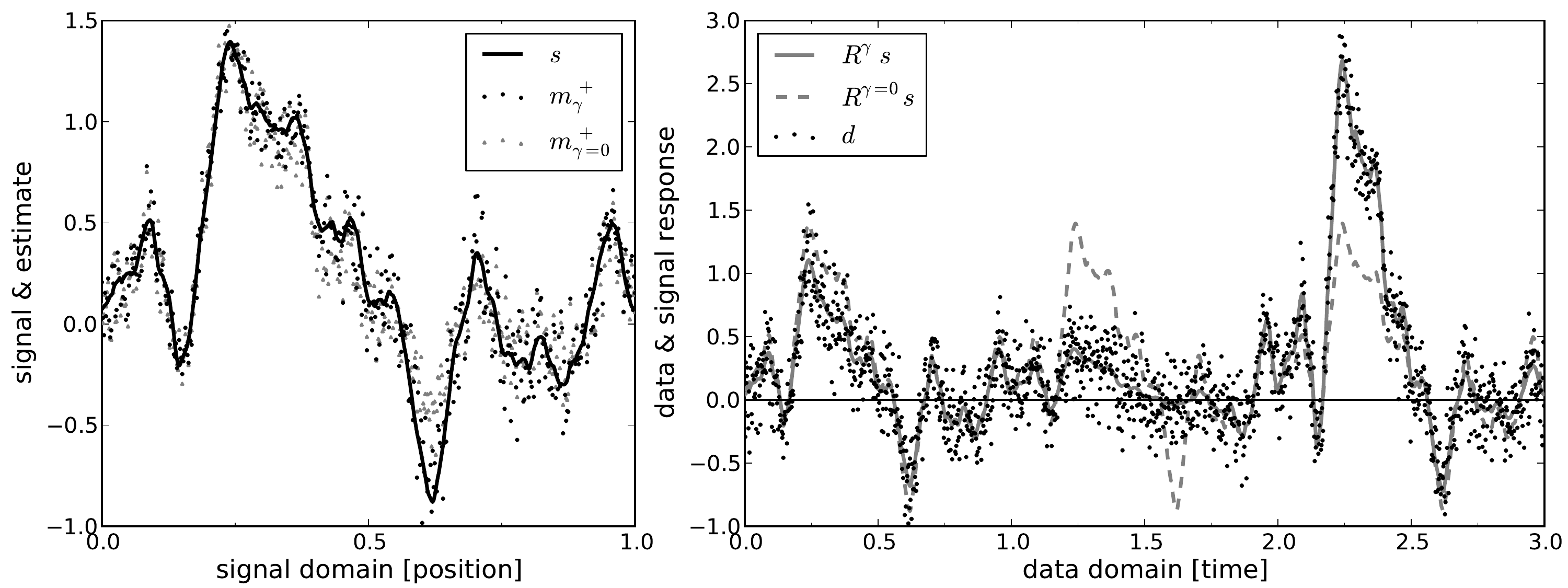}

\caption{\label{fig:Simulated-signal-and}Simulated signal (according to Eqs.
\eqref{eq:signal_prior} and \eqref{eq:spectra-example}) and data
realization from observing the signal three times (according to Eqs.~\eqref{eq:data_model},
\eqref{eq:likelihood} and \eqref{eq:response_example}). Left: Signal
(line), its prior-free and therefore noisy estimation by Eq.~\eqref{eq:signal estimator}
using the correct calibration ($\gamma$, black dots), and $ $no
calibration ($\gamma=0$, gray triangles). Right: Data (dots), the
signal response $R^{\gamma}s$ (gray line, Eqs.~\eqref{eq:response_example}
and \eqref{eq:spectra-example}) and the response in case of zero
calibration, $R^{\gamma=0}s$ (dashed, gray line, basically three
repetitions of the signal pattern). The difference between these two
lines contains information on the calibration. The corresponding gain
curve is shown in Fig.~\eqref{fig:Signal-reconstruction-and}. }
\end{figure*}

A more rigorous and more abstract treatment of the calibration problem
should be addressed now. The signal and data domain are not necessarily
the same anymore as signals live typically in continuous domains (time,
position, or spectral spaces) and data sets are always finite. For
dealing with probabilities over spaces of continuous functions (fields
in physical language) we use the formalism of information field theory
\citep{1999physics..12005L,2009PhRvD..80j5005E,2013AIPC.1553..184E}.

A generic, linear response that maps the signal into data domain will
be assumed. This covers many realistic measurement situations. Unknown
properties of this response are to be calibrated. The unknown signal,
calibration, and noise components are all assumed to fluctuate around
zero with known individual covariances, but no cross-correlations
between them. As we do not assume any higher order statistics of these
components to be known, the maximum entropy criterion \citep{1957PhRv..106..620J,1982ieee...70..939J,2003prth.book.....J,2008arXiv0808.0012C}
suggests we should model our a priori knowledge states as Gaussian
distributions. This does not imply that our analysis is only valid
for Gaussian statistics. If signal, noise, or calibration follow non-Gaussian
distributions and those are known, the here derived methods still
produce sensible results. Just more efficient methods might be constructed
that exploit the additional statistical knowledge. 

We assume that a signal $s=(s_{x})_{x}$ over some continuous domain
(parametrized by $x$) was targeted by a linear measurement device
that produced the finite dimensional data $d=(d_{i})_{i}$ with signal
independent Gaussian noise $n=(n_{i})_{i}$ that includes also calibration
offsets, 
\begin{equation}
d=R\, s+n.\label{eq:data_model}
\end{equation}
The signal response $R=R^{\gamma}=(R_{i\, x}^{\gamma})_{i\, x}$ depends
on the unknown calibration parameters $\gamma=(\gamma_{a})_{a}$ as
well as on the signal and data domain coordinates, here $x$ and $i$,
between which it translates via $\left(R^{\gamma}s\right)_{i}=\int dx\, R_{i\, x}^{\gamma}s_{x}$.
This is the general form for any linear signal response. It not only
embraces the illustrative example of the previous section, where $R_{ix}^{\gamma}=\delta(i-x)\,(1+\gamma_{i}),$
but also a convolution with a calibration dependent kernel, $R_{ix}^{\gamma}=f(i-x,\gamma)$,
Fourier-transformations, $R_{kx}^{\gamma}=\exp(ikx)$, and more complex
measurement situations. 

In general, the response can depend in a very complicated way on the
unknown parameters $\gamma$. We suppose that a first order Taylor
expansion captures the most relevant dependence,
\begin{equation}
R^{\gamma}=R^{0}+\sum_{a}\left.\gamma_{a}R_{,\gamma_{a}}^{\gamma}\right|_{\gamma=0}+\mathcal{O}(\gamma^{2}),
\end{equation}
with $R^{0}=\left.R^{\gamma}\right|_{\gamma=0}$ being the well calibrated
part of the response and $R_{,\gamma_{a}}^{\gamma}=\partial R^{\gamma}/\partial\gamma_{a}$
its linear dependence on the calibration parameter. Thereby, we ignore
second order corrections in $\gamma$. 

To have a compact notation, we define scalar products for the continuous
$u$-dimensional signal domain and its Fourier space as
\begin{equation}
j^{\dagger}s=\int dx^{u}\,\overline{j_{x}}s_{x}=\int\frac{dk^{u}}{(2\pi)^{u}}\,\overline{j_{k}}s_{k},\label{eq:signal space scalar product}
\end{equation}
for the discrete data domain as
\begin{equation}
n^{\dagger}d=\sum_{i}\overline{n_{i}}\, d_{i},\label{eq:data space scalar product}
\end{equation}
and for the calibration parameter domain something analog to \eqref{eq:data space scalar product}
or \eqref{eq:signal space scalar product}, depending on whether the
calibration parameters form a discrete set or a continuous function.
Discrete calibration parameters are instrument gains, since there
are at most a finite number of parameters per data value, so that
the calibration domain can be mapped onto the data domain (the set
of data indices). A continuous set of calibration parameters would
be the spatial sensitivity map of a telescope, the so-called telescope
beam, for which the domain in which the calibration parameters reside
(the sphere $\mathcal{S}^{2}$ of directions in the telescope frame)
can often be mapped onto the signal domain (positions in the sky,
also $\mathcal{S}^{2}$).In order to have an illustrative case, we
assume further that the signal obeys a priori a Gaussian distribution,
\begin{equation}
\mathcal{P}(s)=\mathcal{G}(s,\, S)\equiv\frac{1}{|2\pi S|^{\nicefrac{1}{2}}}\,\exp\left(-\frac{1}{2}\, s^{\dagger}S^{-1}s\right),\label{eq:signal_prior}
\end{equation}
with known covariance $S=\left\langle s\, s^{\dagger}\right\rangle _{(s)}=\int\mathcal{D}s\, s\, s^{\dagger}\,\mathcal{P}(s)$.
This and other covariances are assumed here to be known either from
similar previous measurements or on theoretical grounds. In practice,
they might need to be determined from the data themselves. This is
often well possible, as shown in Refs.~\citep{2011PhRvD..83j5014E,2010PhRvE..82e1112E,2011PhRvE..84d1118O,2013PhRvE..87c2136O},
and explained in Sec.~\ref{sub:Power-spectrum-estimation}. The extension
to non-Gaussian cases can be treated in future studies along the lines
sketched in Refs.~\citep{2009PhRvD..80j5005E,2011PhRvD..83j5014E,2010PhRvE..82e1112E,2011PhRvE..84d1118O}.

The noise covariance $N=\left\langle n\, n^{\dagger}\right\rangle _{(n)}$
is assumed to be known as well, leading to the likelihood 
\begin{equation}
\mathcal{P}(d|s,\,\gamma)=\mathcal{P}(n=d-R^{\gamma}s|s)=\mathcal{G}(d-R^{\gamma}s,\, N).\label{eq:likelihood}
\end{equation}

Likelihood and prior can be combined into the joint probability of
data and signal, $\mathcal{P}(d,\, s|\gamma)=\mathcal{P}(d|s,\,\gamma)\,\mathcal{P}(s|\gamma)=\mathcal{P}(d|s,\,\gamma)\,\mathcal{P}(s)$
(see Eq.~\eqref{eq:independence} for the last step), from which
the signal posterior for known calibration can be obtained via Bayes
theorem,
\[
\mathcal{P}(s|d,\,\gamma)=\frac{\mathcal{P}(d,\, s|\gamma)}{\mathcal{P}(d|\gamma)}=\frac{e^{-\mathcal{H}(d,\, s|\gamma)}}{\mathcal{Z}(d|\gamma)}.
\]
Here, we have introduced the information Hamiltonian, $ $$\mathcal{H}(d,\, s|\gamma)\equiv-\log\,\mathcal{P}(d,\, s|\gamma)$,
and its partition function,
\begin{equation}
\mathcal{Z}(d|\gamma)\equiv\int\mathcal{D}s\, e^{-\mathcal{H}(d,\, s|\gamma)}=\int\mathcal{D}s\,\mathcal{P}(d,\, s|\gamma)=\mathcal{P}(d|\gamma),
\end{equation}
in order to exploit the mathematical and conceptual analogies of Bayesian
inference and thermodynamics.

\subsection{Wiener filter}

\begin{figure*}
\includegraphics[width=1\textwidth]{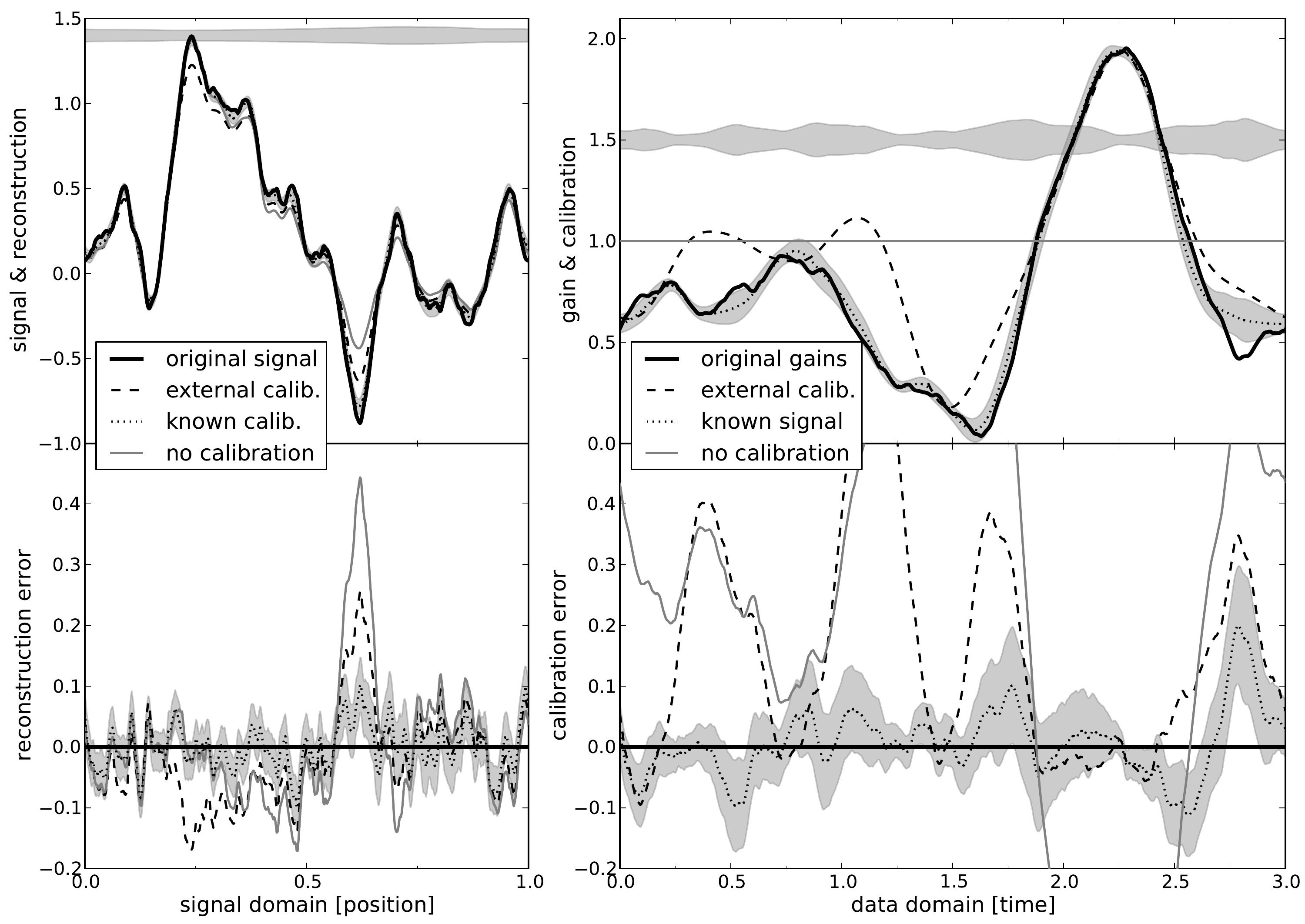}\caption{\label{fig:Signal-reconstruction-and}Signal reconstruction and calibration
without \emph{selfcal}. Left: Original signal (as in Fig.~\eqref{fig:Simulated-signal-and}),
and signal reconstructions using only the external calibration data
(according to Eq.~\eqref{eq:gamma*} with external calibrator $c$
at only four moments as described in Sec.~\ref{sec:Numerical-example}),
the correct gains ($g=1+\gamma$, Eq.~\eqref{eq:Wiener filter}),
and no calibration ($g=1$, $\gamma=0$ in Eq.~\eqref{eq:Wiener filter}).
Right: Original gains (according to Eqs.~\eqref{eq:gain-prior} and
\eqref{eq:spectra-example}) and their calibration reconstruction
using only the external calibration data (as on the right hand side),
calibrating on the correct signal (Eq.~\eqref{eq:gamma*} with $c=s$),
and assuming no calibration ($g=1$, $\gamma=0$). The gray areas
in the left and right panels show the one sigma posterior uncertainties
of the signal and calibration reconstructions using the correct calibration
and signal, respectively. These are the accuracies of the best achievable
reconstructions and show that recovering the calibration accurately
is more difficult than recovering the signal. In the top panels these
uncertainties are shown twice, once around the signal/calibration
reconstructions and once at an arbitrary location for better visual
inspection of their structures.}
\end{figure*}
\begin{figure*}
\includegraphics[width=1\textwidth]{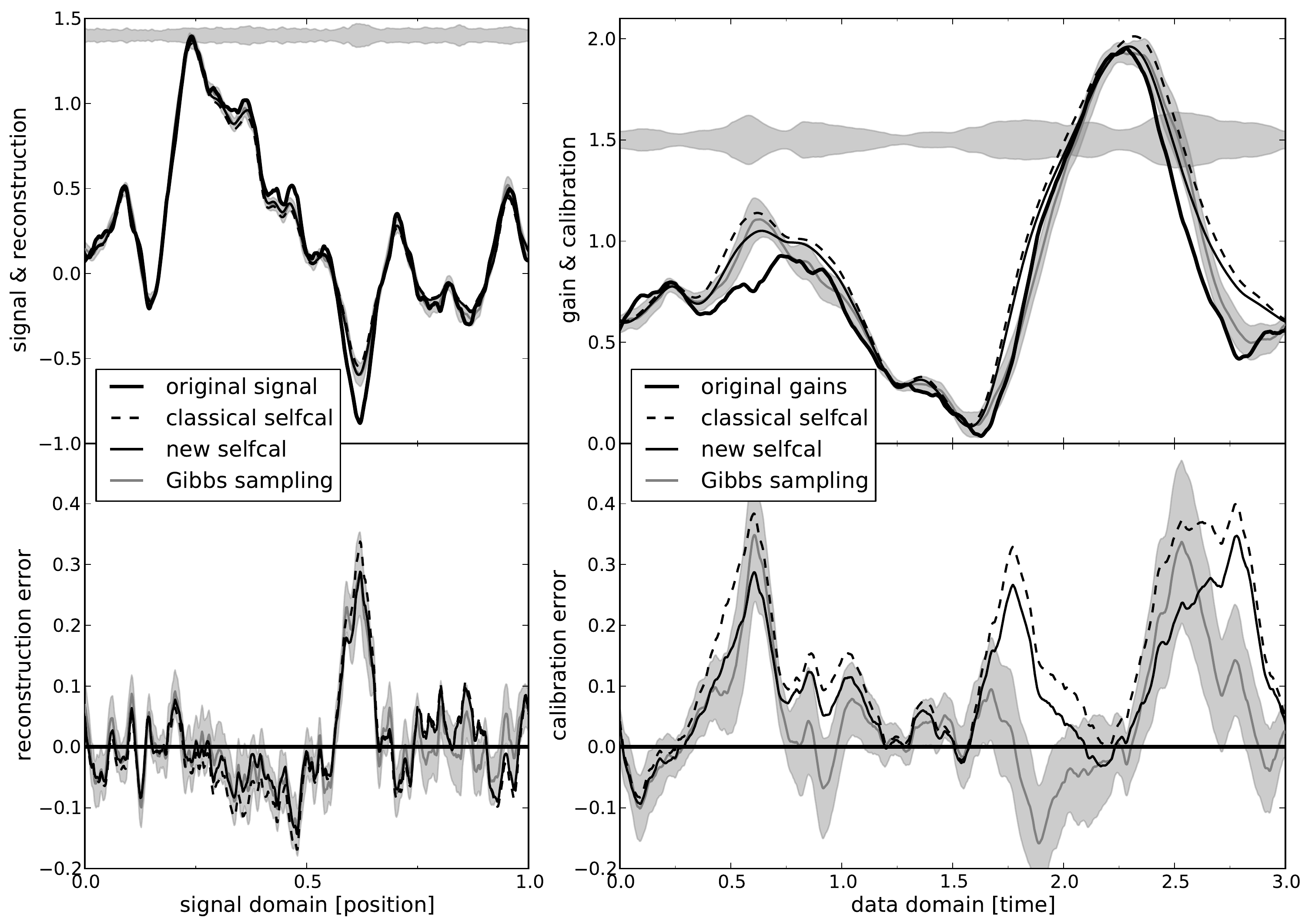}\caption{\label{fig:Signal-reconstruction-with_selfcal}Signal reconstruction
and calibration using \emph{selfcal}. Left: Signal (as in previous
figures) and its reconstruction using \emph{classical} \emph{selfcal}
(iterating Eq.~\eqref{eq:Wiener filter} with $\gamma=\gamma^{*}$
to get $m$ and Eq.~\eqref{eq:gamma*} with $c=m$ to get $\gamma^{*}$;
precisely Eqs.~\eqref{eq:selcal} are used with $T=0$),\emph{ new}
\emph{selfcal} (Eqs.~\eqref{eq:selcal} with $T=1$), and using Gibbs
sampling (see Sec.~\eqref{sub:Gibbs-smapling}). Right: The gain
curve and its reconstructions using the \emph{classical} (Eqs.~\eqref{eq:selcal}
with $T=0$) and the \emph{new selfcal} (Eqs.~\eqref{eq:selcal}
with $T=1$) scheme as well as using Gibbs sampling (Sec.~\eqref{sub:Gibbs-smapling}).
The uncertainty estimates of the Gibbs sampling are shown as gray
bands in all panels. In the top panels it is shown twice, once around
the Gibbs sampling mean and once at an arbitrary location for better
visual inspection of its structure.}
\end{figure*}

Under these conditions, the optimal signal reconstruction for a given
calibration $\gamma$ is known to be the Wiener filter (e.g., see
Ref.~\citep{2009PhRvD..80j5005E}),
\begin{equation}
m^{\gamma}=\langle s\rangle_{(s|d\,\gamma)}=D^{\gamma}\, j^{\gamma},\label{eq:Wiener filter}
\end{equation}
where 
\begin{eqnarray}
D^{\gamma} & = & \left(S^{-1}+R^{\gamma\dagger}N^{-1}R^{\gamma}\right)^{-1},\label{eq:D}\\
j^{\gamma} & = & R^{\gamma\dagger}N^{-1}d,
\end{eqnarray}
are the information propagator (or Wiener variance) and information
source, respectively \citep{2009PhRvD..80j5005E}. This formula is
equivalent to the data space centric formula for Wiener filtering,
see Eq.~\eqref{eq:Wiener filter data space}, we had argued to be
the optimal linear filter (minimizing Eq.~\eqref{eq:square error}).
The remaining a posteriori uncertainty of the signal is given by the
Wiener variance,

\begin{equation}
\left\langle (s-m^{\gamma})\,(s-m^{\gamma})^{\dagger}\right\rangle _{(s|d,\,\gamma)}=D^{\gamma}.
\end{equation}

Since the signal posterior for known calibration is a Gaussian (for
this case composed of a Gaussian prior and likelihood, and a linear
response), it must be
\begin{equation}
\mathcal{P}(s|d,\,\gamma)=\mathcal{G}(s-m^{\gamma},D^{\gamma}),\label{eq:signal posterior known calib}
\end{equation}
as can also be verified by a direct calculation.

The often used so-called \textit{prior-free} or maximum likelihood
reconstruction can as well be reproduced by taking the limit of $S\rightarrow\infty$
or $S^{-1}\rightarrow0$, which removes any prior contribution to
the filter formula, and interpreting the matrix inversion in Eq.~\eqref{eq:D}
as a pseudo-inverse%
\footnote{We define the pseudo-inverse of a Hermetian matrix $A=\sum_{i}a_{i}a_{i}^{\dagger}\lambda_{i}$
with eigenvalues $\lambda_{i}$ and normalized eingenvectors $a_{i}$
as 
\[
A^{+}=\sum_{i}a_{i}a_{i}^{\dagger}\begin{cases}
\lambda_{i}^{-1} & \lambda_{i}\neq0\\
0 & \lambda_{i}=0
\end{cases}.
\]
}, so that 
\begin{equation}
m_{\gamma}^{+}=(R^{\gamma\dagger}N^{-1}R^{\gamma})^{+}R^{\gamma\dagger}N^{-1}d.\label{eq:signal estimator}
\end{equation}
This prior-free signal estimator is very noisy, as can be seen from
Fig.~\ref{fig:Simulated-signal-and}. There, a simulated signal,
the resulting data, and the corresponding prior-free signal estimator
are shown. The latter exhibits a lot of noise%
\footnote{This noise could be reduced by binning, averaging, or smoothing. This
requires that an averaging length scale has to be specified. The optimal
averaging length scale should be a trade off between suppressing noise
and keeping signal features. However, the Wiener filter, see Eq.~\eqref{eq:Wiener filter},
performs already this averaging in an optimal way (minimizing Eq.~\eqref{eq:square error})
with an averaging length that depends on the local signal-response-to-noise
ratio and therefore can vary with position. We therefore use in the
following the Wiener filter method and regard binning and averaging
scheme applied in practice as approximative realizations thereof. %
} compared to the reconstructions exploiting the knowledge on covariances
shown in Fig.~\ref{fig:Signal-reconstruction-and}.

In the following, we often suppress the $\gamma$-dependence of $R$,
$D$, $j$, $m$ and other quantities for notational compactness,
as we also do not note explicitly that $m$ is a function of the data
$d$.

\subsection{Power spectrum estimation\label{sub:Power-spectrum-estimation}}

A problem in setting up the Wiener filter is often that the signal
and noise covariances are not known precisely or might even be completely
unknown. Thus, these need to be inferred from the same data used for
imaging. The proper way is to formulate hyper-priors on these spectra,
and to solve the combined problem of simultaneous signal and spectra
recovery. A suitable, however numerically expensive method for this
is Gibbs sampling \citep{gibbsamp,2004PhRvD..70h3511W}, here introduced
in Sec. \ref{sub:Gibbs-smapling}. 

An approximative, but numerically cheaper approach of iteratively
analyzing a reconstruction for its covariance and using this covariance
in improved reconstructions was developed in \citep{2011PhRvD..83j5014E,2010PhRvE..82e1112E,2011PhRvE..84d1118O,2013PhRvE..87c2136O}.
The basic idea works for a statistical stationary signal, for which
the signal covariance is diagonal in Fourier space, with the power
spectrum $P_{s}(k)=\langle|s_{k}|^{2}\rangle_{(s)}$ on the diagonal
(here, quantities with the index $k$ denote Fourier transformed quantities
like $s_{k}=\int dx\, s_{x}\exp(ikx)$). In case of a Jeffreys prior
on the power spectra, a uniform distribution on logarithmic scale,
the formula to get a point estimate for the spectrum is

\begin{equation}
P_{s}(k)\approx|m_{k}|^{2}+D_{kk},\label{eq:spectrum-estimation}
\end{equation}
where $D_{kk}$ corrects for the missing variance in the Wiener filter
reconstruction $m$. Eq.~\eqref{eq:Wiener filter} for $m$, Eq.
\eqref{eq:D} for $D$, and Eq. \eqref{eq:spectrum-estimation} for
$P_{s}(k)$ have to be iterated until convergence. The accuracy of
this spectral estimate can be improved by averaging Fourier modes
with similar spectrum and by exploiting available prior information
on the spectral values and their smoothness as a function of wavevector
\citep{2010arXiv1002.2928E,2013PhRvE..87c2136O}. The method can even
be extended to estimate simultaneously the signal and noise covariance
\citep{2011PhRvE..84d1118O} and be combined with non-linear signal
estimators \citep{2010PhRvE..82e1112E,2013arXiv1311.1888S,2013arXiv1311.5282J}.
Thus, an unknown covariance can be dealt with in principle. In order
to be able to concentrate on the essentials of the calibration problem,
we assume in the following known covariances as well as Gaussian prior
distributions for signal, noise, and unknown calibration parameters.

Although we showed that methods exist to obtain estimates of these
covariances from the data themselves, we should investigate how sensitive
a reconstruction is to inaccuracies in those estimates. For this,
we consider the special case of a signal and data space being identical,
the response the identity matrix, and signal and noise being statistically
homogeneous processes. In this case, their covariances are diagonal
in Fourier space, with the corresponding power spectra $P_{s}(k)$
and $P_{n}(k)$ on the diagonals. The Wiener filter in Fourier space
is then

\begin{equation}
m_{k}=\frac{d_{k}}{1+P_{n}(k)/P_{s}(k)}.
\end{equation}
Thus, high signal-to-noise (S/N) modes with $P_{n}(k)/P_{s}(k)\ll1$
are unmodified by the filter, $m_{k}\approx d_{k}$, whereas low S/N
modes with $P_{n}(k)/P_{s}(k)\gg1$ are strongly suppressed by the
filter, $m_{k}\rightarrow0$. Only for S/N ratios around one, the
precise value of the spectra matters. Over- or underestimation of
the S/N ratio leads to too much noise in the reconstruction or an
unnecessary strong signal suppression, respectively. However, this
effect is mainly relevant for the modes with a S/N around unity. Therefore,
moderate inaccuracies in the power spectra or covariances lead only
to a minor degradation of the reconstruction fidelity \citep{2011PhRvD..83j5014E}.
This usually also holds for more complex measurement situations than
used in this argumentation \citep{2012A&A...542A..93O}.

\subsection{External calibration}

Somehow, the calibration parameters $\gamma$ have to be measured
in order that Eq.~\eqref{eq:Wiener filter} can be used to determine
the a posteriori mean of the signal. The simplest strategy is to use
an external calibrator signal as a known reference from which the
calibration can be determined.

In case the calibration parameters are constant in time, they can
be determined using a known calibration signal $c$ and then be transferred
to the measurement of interest. The calibrator $c=(c_{x})_{x}$ is
just a signal, which ideally is known before the measurement, which
is strong enough to have a signal response $R^{\gamma}\, c$ dominating
over the noise, and which is sufficiently complex to probe the relevant
calibration uncertainties. The last requirement means that the calibrator
response should depend on the calibration parameters such that $\partial R^{\gamma}\, c/\partial\gamma_{a}\equiv R_{,\gamma_{a}}^{\gamma}c$
is significantly non-zero for any relevant $\gamma_{a}$. 

For the calibration parameters we assume here and in the following
a Gaussian, zero-centered prior%
\footnote{The mean can always be subtracted by a redefinition of the calibration
parameters. As it is known, it should be part of the known part of
the response, whereas the calibration parameter should only affect
the unknown part. %
},
\begin{equation}
\mathcal{P}(\gamma)=\mathcal{G}(\gamma,\,\Gamma),\label{eq:gain-prior}
\end{equation}
with known uncertainty covariance $\Gamma=\langle\gamma\,\gamma^{\dagger}\rangle_{(\gamma)}$.
The knowledge on $\Gamma$ can come from theoretical considerations,
previous measurements, or might be obtained from the data themselves.
Prior and likelihood, Eq. \eqref{eq:likelihood}, form the joint probability
$\mathcal{P}(d,\gamma|\, c)=\mathcal{P}(d|\gamma,\, c)\,\mathcal{P}(\gamma)$
that contains all available information on the calibration. 

In general, the calibration inference from this is a non-linear and
non-trivial problem. In many cases, the MAP approximation provides
a reasonable estimate for $\gamma$. This is obtained by minimizing
the corresponding Hamiltonian 
\begin{eqnarray}
\mathcal{H}(d,\,\gamma|c) & = & -\log\mathcal{P}(d,\,\gamma|c)\nonumber \\
 & = & \frac{1}{2}\gamma^{\dagger}\Gamma\gamma+\frac{1}{2}\left(d-R^{\gamma}c\right)^{\dagger}N^{-1}\left(d-R^{\gamma}c\right)\nonumber \\
 &  & +\mathrm{const}.
\end{eqnarray}
The gradient of this Hamiltonian, 

\begin{equation}
\frac{\partial\mathcal{H}(d,\,\gamma|c)}{\partial\gamma}=\Gamma^{-1}\gamma-c^{\dagger}R_{,\gamma}^{\gamma\dagger}N^{-1}(d-R\, c),\label{eq:externalMAPcalib}
\end{equation}
should then be followed (downhill) until it is zero and the Hamiltonian
minimal. Here $\left(R_{,\gamma}^{\gamma}\right)_{a}\equiv R_{,a}\equiv\partial R^{\gamma}/\partial\gamma_{a}$
denotes the derivative of the response with respect to the calibration.
It is apparent that the discrepancy of the data from the calibration
signal response, $d-R^{\gamma}c$ , drives the calibration solution
away from the default value $\gamma=0$ preferred by the prior of
the calibration, $\mathcal{P}(\gamma)=\mathcal{G}(\gamma,\,\Gamma)$. 

In case the calibration parameters enter only linearly,

\begin{equation}
R^{\gamma}=\R^{0}+\sum_{a}\gamma_{a}\R^{a},\label{eq:linearCalibration}
\end{equation}
with $\R^{0}$ and $\R^{a}$ known and $\gamma$-independent, we have
$R_{,a}=R_{,\gamma_{a}}^{\gamma}=\R^{a}$ and the minimum of the Hamiltonian
is at
\begin{eqnarray}
\gamma^{\star} & = & \Delta\, h,\mbox{ with}\label{eq:gamma*}\\
\Delta_{ab}^{-1} & = & \Gamma_{ab}^{-1}+c^{\dagger}\R^{a\dagger}N^{-1}\R^{b}c\mbox{, and}\nonumber \\
h_{b} & = & c^{\dagger}\R^{b\dagger}N^{-1}\left(d-\R^{0}c\right).\nonumber 
\end{eqnarray}
This MAPstimator for the calibration $\gamma^{\star}$ is actually
also the calibration posterior mean $\langle\gamma\rangle_{(\gamma|d,\, c)}$,
since this particular posterior is a Gaussian for which mean and maximum
coincide. This Gaussian calibration posterior is
\begin{equation}
\mathcal{P}(\gamma|d,\, c)=\mathcal{G}(\gamma-\gamma^{\star},\Delta),\label{eq:externCalibPost}
\end{equation}
with the uncertainty covariance $\Delta=\langle(\gamma-\gamma^{\star})\,(\gamma-\gamma^{\star})^{\dagger}\rangle_{(\gamma|d,\, c)}$
given in Eq.~\eqref{eq:gamma*}. 

In this specific linear calibration case, external calibration is
Wiener filtering. This can be seen by comparing Eq.~\eqref{eq:gamma*}
with the Wiener filter equations for the signal, Eqs.~\eqref{eq:Wiener filter}-\eqref{eq:D},
while recognizing that the roles of the following terms correspond
to each other: $\gamma^{\star}\leftrightarrow m$, $\Gamma\leftrightarrow S$,
$N\leftrightarrow N$, $ $ $\R^{0}c\leftrightarrow R$, $\Delta\leftrightarrow D$,
$ $$d-\R^{0}c\leftrightarrow d$, and $h\leftrightarrow j$$ $. 

There is, however, an interesting difference. The signal information
source $j=R^{\dagger}N^{-1}d\approx R^{\dagger}N^{-1}R\, s=\left(B^{0}N^{-1}B^{0}+\mathcal{O}(\gamma)\right)\, s$
contains a calibration-independent term, $B^{0}N^{-1}B^{0}\, s$,
which reacts to $s$ even when $\gamma=0$, whereas the calibration
information source $h_{a}\approx c^{\dagger}\R^{a\dagger}N^{-1}\left(R-\R^{0}\right)\, c=\sum_{b}\left(c^{\dagger}\R^{a\dagger}N^{-1}\R^{b}\, c\right)\,\gamma_{b}=\sum_{a}Q^{ab}\,\gamma_{b}$
is a quadratic function of the calibration signal $c$, which vanishes
for locations with vanishing $c$. The quadratic dependence of $Q^{ab}=c^{\dagger}\R^{a\dagger}N^{-1}\R^{b}\, c$
on the calibration signal strength will become important again later
on, when we investigate \emph{selfcal}, the attempt to calibrate on
an unknown signal.

\subsection{Calibration binning}

It should be noted that the usage of an a priori calibration covariance
$\Gamma=\langle\gamma\gamma^{\dagger}\rangle_{(\gamma)}$$ $ to suppress
the calibration estimation noise is not standard practice. Instead,
bin-averaging and interpolation is often performed on $\chi^{2}$
or maximum likelihood calibration estimators. 

There is, however, no consensus on the question how to choose the
bin size and interpolation scheme. The optimal bin size should, on
the one hand, be sufficiently large to average down the noise, and
on the other hand, be sufficiently small in order not to iron out
existing small scale (spatial or temporal) variations in the gain
parameters. Therefore, the optimal bin choice depends on the interplay
of expected calibration variations as encoded in $\Gamma$, the noise
level $N$, and the strength of the calibrator signal in data space
$R^{\gamma}\, c$. Since all these elements are part of the MAP gain
estimator, c.f.~Eq.~\eqref{eq:externalMAPcalib}, that reduces to
the Wiener filter solution for linear calibration problems, Eq.~\eqref{eq:gamma*},
we expect the latter to implement (nearly) an optimal averaging and
interpolation scheme. The optimal bin size could be read of from this
scheme (it should be of the order of the correlation length of $\Delta$),
or, even better, the binning and averaging be replaced with the more
accurate calibration solution given by Eq.~\eqref{eq:externalMAPcalib}
or Eq.~\eqref{eq:gamma*}. 

In the following, we use the un-binned, non-parametric Wiener filter
solution since it is optimal or close to optimal. We believe that
binning schemes used in practice and chosen with experience can come
sufficiently close to the Wiener filter performance as that the difference
does not matter much for our discussion. When they matter, an adoption
of the here proposed non-parametric Wiener filter calibration methodology
would be beneficial and highly recommended for the application.

\subsection{Gibbs sampling\label{sub:Gibbs-smapling} }

The signal and the calibration are the two unknowns. Their joint posterior
probability distribution $\mathcal{P}(s,\gamma|d)$ can be probed
via Gibbs sampling in case it is possible to draw samples from $\mathcal{P}(s|d,\gamma)$
and $\mathcal{P}(\gamma|d,s)$ \citep{2004astro.ph..1623W,2004PhRvD..70h3511W,2010MNRAS.406...60J}.
These are Gaussian distributions in our case, given by Eqs.~\eqref{eq:signal posterior known calib}
and \eqref{eq:externCalibPost} (with $c=s$), respectively, from
which it is well possible to draw samples. The Gibbs sampling procedure
is then to update a combined signal and calibration probe $p^{(i)}=(s^{(i)},\gamma^{(i)})\rightarrow p^{(i+1)}=(s^{(i+1)},\gamma^{(i+1)})$
via
\begin{eqnarray}
s^{(i+1)} & \hookleftarrow & \mathcal{P}(s^{(i+1)}|d,\gamma^{(i)}),\nonumber \\
\gamma^{(i+1)} & \hookleftarrow & \mathcal{P}(\gamma^{(i+1)}|d,s^{(i+1)}).
\end{eqnarray}
If this updating is an ergodic process for the combined $p$-space,
as it is in our case of Gaussian probabilities, the sample distribution
can be shown to converge towards $\mathcal{P}(s,\gamma|d)$.

Marginalization with respect to $s$ or $\gamma$ to obtain $\mathcal{P}(\gamma|d)$
and $\mathcal{P}(s|d)$, respectively, can be obtained from the samples
by forgetting the corresponding marginal variable. Any posterior average,
like $\langle s\rangle_{(s|d)}$, is given by the corresponding sample
averages. The Gibbs sampling provides therefore a route to calculate
any desired estimate from the full posterior, without invoking approximations,
except for replacing the posterior integration by finite sampling
and therefore getting some shot noise. Beating down this shot noise
by generating a large number of samples can become computationally
expensive, why it makes sense also to investigate analytical alternatives
as we do in the following. Analytical investigations also provide
deeper insight into the structure of the problem, which is less easily
obtained from the sampling machinery.

Anyhow, we provide Gibbs sampling results as an optimal benchmark
for the different \emph{selfcal} schemes implemented in Sec. \ref{sec:Numerical-example}.

\subsection{Calibration marginalized imaging\label{sub:Calibration-marginalized-imaging}}

All relevant information on the signal is contained in the calibration
marginalized posterior,
\begin{eqnarray}
\mathcal{P}(s|d) & = & \int\mathcal{D}\gamma\, P(s,\gamma|d).
\end{eqnarray}
In the case of linear calibration coefficients, $R^{\gamma}=\R^{0}+\sum_{a}\gamma_{a}\R^{a}$,
see Eq.~\eqref{eq:linearCalibration}, the calibration marginalized
likelihood, from which this posterior can be constructed, can be calculated
analytically,
\begin{eqnarray}
\mathcal{P}(d|s) & = & \int\mathcal{D}\gamma\,\mathcal{P}(d|s,\gamma)\mathcal{\, P}(\gamma)\nonumber \\
 & = & \int\mathcal{D}\gamma\,\mathcal{G}(d-R^{\gamma}s,N)\mathcal{\, G}(\gamma,\Gamma)\label{eq:marginal likelihood}\\
 & = & \mathcal{G}\left(d-\R^{0}s,N+\sum_{ab}\R^{a}s\Gamma_{ab}s^{\dagger}\R^{b\dagger}\right).\nonumber 
\end{eqnarray}
This result can be found in Ref.~\citep{2002MNRAS.335.1193B}.%
\footnote{One can also simply calculate the first two moments of the data given
the signal averaged over noise and calibration realizations, $\bar{d}=\langle d\rangle_{(n,\gamma|s)}=\R^{0}s$,
$\langle(d-\bar{d})\,(d-\bar{d})^{\dagger}\rangle_{(n,\gamma|s)}=N+\sum_{ab}\R^{a}s\Gamma_{ab}s^{\dagger}\R^{b\dagger}$,
and realize that the calibration marginalized likelihood has to be
a Gaussian with this mean and variance, since both, noise and calibration
uncertainty, just add Gaussian variance to the data.%
}

The resulting posterior $\mathcal{P}(s|d)=\mathcal{P}(d|s)\,\mathcal{P}(s)/\mathcal{P}(d)$
is non-Gaussian, as the signal field appears as part of the calibration
marginalized effective noise, $N+\sum_{ab}\R^{a}s\Gamma_{ab}s^{\dagger}\R^{b\dagger}$.
Ideally, the mean of this signal posterior is calculated since this
gives the optimal signal estimate. 

However, integrating over this non-Gaussian function is often infeasible.
The quadratic dependence of the effective noise on the unknown signal
inhibits that this can be calculated via a simple Gaussian integration.
In high-dimensional settings, Monte-Carlo methods used to estimate
phase space integrals might become too expensive. In such cases, approximative
strategies are needed. One is to use the MAP estimator for this posterior.
However, due to the skewness of the distribution, this can be expected
to give biased results. It is better to characterize the calibration
posterior $\mathcal{P}(\gamma|d)$ by its mean $\gamma^{\star}$ and
uncertainty covariance $\Delta$ and to use them to construct an approximative
signal estimation.

Let us assume we managed somehow to estimate the calibration as $\gamma^{\star}$
with some uncertainty covariance $\Delta$ and that we can well approximate%
\footnote{In the case of an external calibration of only linear calibration
parameters, Eq.~\eqref{eq:linearCalibration}, we had shown in Eq.~\eqref{eq:externCalibPost}
this to be an exact result.%
} 
\begin{equation}
\mathcal{P}(\gamma|d)\approx\mathcal{G}(\gamma-\gamma^{\star},\Delta).\label{eq:Gauss approx. for calib posterior}
\end{equation}

For a Gaussian signal field with $\mathcal{P}(s)=\mathcal{G}(s,\, S)$
we could simply use $\gamma=\gamma^{\star}$ in the Wiener filter
formula, Eq.~\eqref{eq:Wiener filter}. However, this is suboptimal
if the calibration uncertainty is significant. In that case, correction
terms might become important, which we calculate now to first order
in the calibration uncertainty $\Delta$.

The optimal, calibration marginalized signal estimator is
\begin{eqnarray}
m & = & \langle s\rangle_{(s|d)}=\left\langle \left\langle s\right\rangle _{(s|d,\,\gamma)}\right\rangle _{(\gamma|d)}\label{eq:calib marginalized signal}\\
 & = & \int\mathcal{D}\gamma\,\mathcal{P}(\gamma|d)\,\left\langle s\right\rangle _{(s|d,\,\gamma)}\nonumber \\
 & \approx & \int\mathcal{D}\gamma\,\mathcal{G}(\gamma-\gamma^{\star},\Delta)\, m^{\gamma}\nonumber \\
 & \approx & D\,\,\left\{ j+\frac{1}{2}\sum_{ab}\left[\Delta_{ab}\,\left(j_{,ba}-M_{,ba}D\, j\right.\right.\right.\!\!\!\!\!\!\!\!\!\!\nonumber \\
 &  & +\left.\left.2\, M_{,b}D\, M_{,a}D\, j-2\, M_{,b}D\, j_{,a}\right)\right]\Biggr\}_{\gamma=\gamma^{\star}}.\!\!\!\!\!\!\!\!\!\!\nonumber 
\end{eqnarray}
In the last step, we Taylor expanded $m^{\gamma}=D^{\gamma}\, j^{\gamma}$
up to second order in $\gamma-\gamma^{\star}$, performed the Gaussian
integration, exploited the Hermitian symmetry of $\Delta$, suppressed
in the notation the dependence of all calibration dependent terms
on $\gamma$, and introduced further the notations 
\begin{eqnarray}
M & = & R^{\dagger}N^{-1}R,\nonumber \\
M_{,a} & = & R_{,\gamma_{a}}^{\dagger}N^{-1}R+R^{\dagger}N^{-1}R_{,\gamma_{a}},\nonumber \\
M_{,ab} & = & R_{,\gamma_{a}}^{\dagger}N^{-1}R_{,\gamma_{b}}+R_{,\gamma_{b}}^{\dagger}N^{-1}R_{,\gamma_{a}}\nonumber \\
 &  & +R_{,\gamma_{a}\gamma_{b}}^{\dagger}N^{-1}R+R^{\dagger}N^{-1}R_{,\gamma_{a}\gamma_{b}},\nonumber \\
j & = & R^{\dagger}N^{-1}d,\nonumber \\
j_{,a} & = & R_{,\gamma_{a}}^{\dagger}N^{-1}d,\mbox{ and}\nonumber \\
j_{,ab} & = & R_{,\gamma_{a}\gamma_{b}}^{\dagger}N^{-1}d.
\end{eqnarray}
In case of only linear calibration parameters as in Eq.~\eqref{eq:linearCalibration},
$R=\R^{0}+\sum_{a}\gamma_{a}\R^{a}$ , the derivatives simplify to
\begin{eqnarray}
M_{,a} & = & \R^{a\dagger}N^{-1}R+R^{\dagger}N^{-1}\R^{a},\\
M_{,ab} & = & \R^{a\dagger}N^{-1}\R^{b}+\R^{b\dagger}N^{-1}\R^{a},\nonumber \\
j_{,a} & = & \R^{a\dagger}N^{-1}d,\mbox{ and }j_{,ab}=0.\nonumber 
\end{eqnarray}

From Eq.~\eqref{eq:calib marginalized signal} it becomes apparent
that the optimal signal reconstruction in the presence of calibration
uncertainties should contain a correction to $m^{\star}=D^{\gamma^{\star}}\, j^{\gamma^{\star}}$,
which corrects for the possibility that certain structures in the
data might well be due to a miscalibration rather than being caused
by real signal structures. Thus, the reconstruction will be less prone
to over-fitting calibration errors. 

The expected level of this correction can, however, be expected to
be moderate in typical situations. The individual correction terms
in Eq.~\eqref{eq:calib marginalized signal} can be paired into similar
ones with opposite signs which partly, but not fully, balance each
other. As a consequence, we expect only a moderate net correction
by them. For the sake of clarity of the following discussion, we will
therefore neglect these corrections and only work with the lowest
order signal estimator $m^{\star}=D^{\gamma^{\star}}\, j^{\gamma^{\star}}$.
The accuracy of this depends, however, crucially on the quality of
the calibration, which should therefore be our focus.

\subsection{Self-calibration}

\subsubsection{Motivation}

In many situations, only insufficient external calibration measurements
are available. In this case, the signal $s$ of scientific interest
has also to serve as a calibration signal. Some \emph{selfcal} procedure
has to be applied in which signal and calibration parameters have
to be determined simultaneously from the same data. 

Furthermore, the case of a perfectly known external calibration is
rarely met in practice. Usually, the calibration signal $c$ was measured
with another imperfect reference instrument as well as with the scientific
instrument that is also used to observe the science signal $s$. We
can now regard the combined measurements ($c$ with reference instrument,
$c$ with scientific instrument, and $s$ with scientific instrument)
as a single measurement, with combined signals, responses, noises,
and calibration parameter sets.

In our mathematical description, we can combine these individual measurements
into a single measurement of a multicomponent signal $s'=(c,\, s)^{\mathrm{t}}$
by a multicomponent instrument delivering the combined data $d'=(d_{c}^{\mathrm{r}},\, d_{c}^{\mathrm{s}},\, d_{s}^{\mathrm{s}})^{\mathrm{t}}$.
Here, the data $d_{c}^{\mathrm{r}}$ result from the measurement of
the calibration signal $c$ with the reference instrument r, the data
$d_{c}^{\mathrm{s}}$ from the calibration measurement of $c$ with
the scientific instrument s, and data $d_{s}^{\mathrm{s}}$ from the
science signal $s$ measurement with the scientific instrument s.
The combined measurement equation reads
\[
\begin{pmatrix}d_{c}^{\mathrm{r}}\\
d_{c}^{\mathrm{s}}\\
d_{s}^{\mathrm{s}}
\end{pmatrix}=\begin{pmatrix}R_{c}^{\mathrm{r}} & 0\\
R_{c}^{\mathrm{s}} & 0\\
0 & R_{s}^{\mathrm{s}}
\end{pmatrix}\begin{pmatrix}c\\
s
\end{pmatrix}+\begin{pmatrix}n_{c}^{\mathrm{r}}\\
n_{c}^{\mathrm{s}}\\
n_{s}^{\mathrm{s}}
\end{pmatrix}\mbox{ or }d'=R'\, s'+n',
\]
with the combined noise vector $n'$, and the combined response $R'$
of the three original measurements. In order that the calibration
measurement provides any benefit, the calibration parameters of the
last two measurements with the scientific instrument need to be identical
or at least sufficiently correlated with each other. 

Since for this combined measurement no external calibration exists
(we have incorporated all external measurements), it should as well
be reconstructed with a \emph{selfcal} scheme.

\subsubsection{Practice}

\emph{Selfcal} usually consists of repeatedly reconstructing the signal,
assuming a calibration to be correct, and determining the calibration,
while assuming the signal to be given. These steps are repeated until
signal and calibration estimates have converged sufficiently. However,
a proof that this converges and the meaning of the fix point seem
are often missing in the \emph{selfcal} literature. 

Using simultaneously MAP estimators for the signal inference and the
calibration actually means that the joint posterior of signal and
calibration parameters is extremized in both unknowns. This is equivalent
to the minimum of the information Hamiltonian,
\begin{eqnarray}
\mathcal{H}(d,\,\gamma,\, s) & = & -\log\mathcal{P}(d,\,\gamma,\, s)\nonumber \\
 & = & \frac{1}{2}\left(d-R^{\gamma}s\right)^{\dagger}N^{-1}\left(d-R^{\gamma}s\right)\nonumber \\
 &  & +\frac{1}{2}\gamma^{\dagger}\Gamma\gamma+\frac{1}{2}\, s^{\dagger}S^{-1}s+\mathrm{const},
\end{eqnarray}
which is as given by
\begin{eqnarray}
0 & = & \frac{\partial\mathcal{H}(d,\,\gamma,\, s)}{\partial s}=D^{-1}\, s-j\biggl|_{\gamma}\mbox{and}\\
0 & = & \frac{\partial\mathcal{H}(d,\,\gamma,\, s)}{\partial\gamma}=\Gamma^{-1}\gamma-s^{\dagger}R_{,\gamma}^{\dagger}N^{-1}(d-R\, s).\nonumber 
\end{eqnarray}
The resulting formula are identical to the Wiener filter signal reconstruction,
Eq.~\eqref{eq:Wiener filter}, and the calibration on this signal,
Eq.~\eqref{eq:externalMAPcalib}. Thus, the joint MAP \emph{selfcal}
scheme is equivalent or at least similar to the usual practice of
iterating signal and calibration estimation.

It has been noticed, e.g.~by Ref.~\citep{2011PhRvD..83j5014E},
that using a joint MAP solution simultaneously for signal and nuisance
parameters (here the unknown calibration, in \citep{2011PhRvD..83j5014E}
the unknown signal covariance) can be sub-optimal. It is better to
use the signal marginalized posterior to determine the calibration
parameters and then to use the resulting parameters in the signal
reconstruction. This approximation is also known under the term \textit{Empirical
Bayes} \citep[e.g., Ref.][]{2012arXiv1204.1470P}.

\subsection{Signal marginalized calibration}

The signal marginalized Hamiltonian, 
\begin{eqnarray}
\mathcal{H}(d,\,\gamma) & = & -\log\int\mathcal{D}s\,\mathcal{P}(d,\,\gamma,\, s)\\
 & = & \frac{1}{2}\,\left(\gamma^{\dagger}\Gamma^{-1}\gamma-\mathrm{Tr}\,\left(\log D\right)-j^{\dagger}D\, j\right)+\mathrm{const},\nonumber 
\end{eqnarray}
can be minimized with respect to $\gamma$ to find the MAP calibration
solution $\gamma^{\star}$. The gradient and Hessian of this Hamiltonian
are
\begin{eqnarray}
\frac{\partial\mathcal{H}(d,\,\gamma)}{\partial\gamma_{a}} & = & \left(\Gamma^{-1}\gamma\right)_{a}+\mathrm{\frac{1}{2}\, Tr}\left(D\, M_{,a}\right)-j^{\dagger}D\, j_{,a}\nonumber \\
 &  & +\frac{1}{2}\, j^{\dagger}D\, M_{,a}D\, j\mbox{ and }\label{eq:calibration eq}\\
\frac{\partial^{2}\mathcal{H}(d,\,\gamma)}{\partial\gamma_{a}\partial\gamma_{b}} & = & \Gamma_{ab}^{-1}+\frac{1}{2}\,\mathrm{Tr}\left(D\, M_{,ab}-D\, M_{,a}D\, M_{,b}\right)\nonumber \\
 &  & +\frac{1}{2}j^{\dagger}D\, M_{,ab}D\, j\nonumber \\
 &  & +j^{\dagger}D\, M_{,a}D\, j_{,b}+j^{\dagger}D\, M_{,b}D\, j_{,a}\nonumber \\
 &  & -j_{,a}^{\dagger}D\, j_{,b}-j^{\dagger}D\, j_{,ab}\nonumber \\
 &  & -j^{\dagger}D\, M_{,a}D\, M_{,b}D\, j.\label{eq:invHesscalib}
\end{eqnarray}
The Hessian can be used to construct an approximative calibration
uncertainty covariance matrix via 
\begin{equation}
\Delta_{ab}^{-1}\approx\left.\frac{\partial^{2}\mathcal{H}(d,\,\gamma)}{\partial\gamma_{a}\partial\gamma_{b}}\right|_{\gamma=\gamma^{\star}}
\end{equation}
so that a Gaussian approximation of the calibration posterior, Eq.~\eqref{eq:Gauss approx. for calib posterior},
as well as an calibration marginalized signal reconstruction, Eq.~\eqref{eq:calib marginalized signal},
can be obtained.

It is instructive, to compare the classical formula used for external
calibration, Eq.~\eqref{eq:externalMAPcalib}, to the one of \emph{selfcal}
\eqref{eq:calibration eq}. For this we have to identify $m=m^{\gamma}=D^{\gamma}j^{\gamma}$
in Eq.~\eqref{eq:calibration eq}, which reads now 
\begin{eqnarray}
\frac{\partial\mathcal{H}(d,\,\gamma)}{\partial\gamma_{a}} & = & \left(\Gamma^{-1}\gamma\right)_{a}+\mathrm{\frac{1}{2}\, Tr}\left(D\, M_{,a}\right)\nonumber \\
 & - & m^{\dagger}R_{,a}N^{-1}\left(d-R\, m\right),\label{eq:gradient H(d,gamma)}
\end{eqnarray}
with $c$ in Eq.~\eqref{eq:externalMAPcalib}. We see that the only
change is the additional term $\mathrm{\frac{1}{2}\, Tr}\left(D\, M_{,a}\right)$,
which ensures that the signal uncertainty is taken into account in
the calibration.

In case of only linear calibration parameters as in Eq.~\eqref{eq:linearCalibration},
$R^{\gamma}=\R^{0}+\sum_{a}\gamma_{a}\R^{a}$ , a nearly closed calibration
formula can be given,
\begin{eqnarray}
\gamma^{\star} & = & \Delta'\, h,\mbox{ with}\label{eq:asdjfkla}\\
\begin{gathered}\Delta'\end{gathered}
_{ab}^{-1} & = & \Gamma_{ab}^{-1}+\mathrm{Tr}\left[\left(m\, m^{\dagger}+D\right)\R^{a\dagger}N^{-1}\R^{b}\right],\mbox{ and}\nonumber \\
h_{b} & = & m^{\dagger}\R^{b\dagger}N^{-1}d-\mathrm{Tr}\left[\left(m\, m^{\dagger}+D\right)\R^{0\dagger}N^{-1}\R^{b}\right].\nonumber 
\end{eqnarray}
This formula is not exactly closed, since $m=m^{\gamma^{\star}}$
and $D=D^{\gamma^{\star}}$ are still calibration dependent. However,
iterations as performed usually in \emph{selfcal} schemes should converge
to a fix point. In practice, one might prefer to use a gradient scheme
based on Eq.~\eqref{eq:gradient H(d,gamma)} rather than to iterate
Eq.~\eqref{eq:asdjfkla} since the latter contains nested matrix
inversions that are numerically expensive. 

The apparent calibration covariance $\Delta'$ is also not exactly
identical to $\Delta$ obtained from the inverse Hessian, Eq.~\eqref{eq:invHesscalib},
since precisely the calibration dependence of $m$ and $D$ were ignored
in the identification of $\Delta'$. It should, however, be a useful
approximation with lower computationally complexity than Eq.~\eqref{eq:invHesscalib}.

A comparison of the calibration formulas, Eqs.~\eqref{eq:gamma*}
and \eqref{eq:asdjfkla} while identifying $c$ with $m$, reveals
the main effect of the signal marginalization. This inserts an additional
signal uncertainty covariance $D$ wherever a term $m\, m^{\dagger}$
appears. As we had seen in case of the external calibration, the quantity
determining how sensitive the calibration information $h$ reacts
to $\gamma,$ $Q^{ab}=s^{\dagger}\R^{a\dagger}N^{-1}\R^{b}\, s=\mathrm{Tr}\left(s\, s^{\dagger}\R^{a\dagger}N^{-1}\R^{b}\right)$
in $h_{a}\approx\sum_{b}Q^{ab}\gamma_{b}$ (neglecting the noise impact),
depends quadratically on the unknown signal $s$. Using $m\, m^{\dagger}$
as an estimator for the quadratic signal $s\, s^{\dagger}$ underestimates
the variance of the latter, since $m$ is a filtered version of $s$
with less power. The correct a posteriori expectation value for $s\, s^{\dagger}$,

\begin{equation}
\left\langle s\, s^{\dagger}\right\rangle _{(s|d,\gamma)}=m\, m^{\dagger}+D,
\end{equation}
contains the signal uncertainty covariance $D$ in order to correct
for this bias. This is therefore the appropriate term to be used in
$Q^{ab}$. 

The calibration propagator $\Delta$ also gets a similar term $Q^{ab}=\mathrm{Tr}\left(\R^{a\dagger}N^{-1}\R^{b}\left(m\, m^{\dagger}+D\right)\right)$
that ensures that a good guess for the signal variance is used in
the term describing the calibration measurement precision. This additional
positive term due to the $D$ correction in $\Delta^{-1}$ decreases
$\Delta$ and makes therefore the calibration reconstruction $\gamma^{\star}=\Delta h$
less reactive to variations in the data. This prevents an over-calibration
on data features that might be caused by noise. Furthermore, the new
\emph{selfcal} scheme corrects a systematic bias of classical \emph{selfcal}
towards delivering higher calibration values%
\footnote{This is valid in the here discussed case in which the signal is obtained
via noise suppressing filtering, otherwise the bias could even be
opposite in cases, in which noise remnants add spurious variance to
the reconstruction.%
}. We therefore suspect the signal marginalized calibration procedure
to provide a more accurate calibration and signal reconstruction than
the classical joint MAP calibration procedure. Whether this is indeed
the case, we investigate numerically.

\section{Numerical example\label{sec:Numerical-example}}

\subsection{Gain uncertainties}

As an illustrative case to compare the performance of the different
calibration schemes we investigate a simple one-dimensional measurement
problem with gain fluctuations in the spirit of the simplistic example
of Sec. \ref{sec:Illustrative-example}. 

A signal field $s=(s_{x})_{x}$ over the periodic domain $\Omega=\{x\}_{x}=[0,\,1)\subset\mathbb{R}$
is observed $u=3$ times by a scanning instrument. The instrument
has a perfect point like response at scanning location $x_{t}=t\mod1$
at time $t$ but a time varying gain $g_{t}=1+\gamma_{t}$. The instrument
samples with a period $\tau=2^{-9}\approx2\cdot10^{-3}$$ $ so that
the $i$th data point is at location $x_{i\tau}=(i\tau)\mod1$. It
is convenient to regard the data as a function of time (which is discrete
with period $\tau$, so that $t\in\{0,\,\tau,\,2\tau,\,\ldots u\}$)
and to exploit the fact that the spatial and temporal coordinates
are well aligned (except that the temporal domain is $u$ times larger
than the spatial domain). 

The response operator
\begin{equation}
R_{tx}=(1+\gamma_{t})\,\delta(x-x_{t})\label{eq:response_example}
\end{equation}
is of the linear calibration parameter form $R^{\gamma}=\R^{0}+\sum_{a}\gamma_{a}\R^{a}$,
Eq.~\eqref{eq:linearCalibration}, with $\R_{tx}^{0}=\delta_{xx_{t}}$
and $\R_{tx}^{a}=\delta_{at}\delta_{xx_{t}}$ so that $R_{tx,t'}=\left(R_{tx}^{\gamma}\right)_{,\gamma_{t'}}=\delta_{tt'}\delta_{xx_{t}}$
and $R_{tx,t't''}=\left(R_{tx}\right)_{,\gamma_{t'}\gamma_{t''}}=0$.%
\footnote{As a consequence of this simple response and noise structure while
assuming white noise with $N_{tt'}=\sigma_{n}^{2}\delta_{tt'}$, we
get 
\begin{eqnarray*}
M_{xy} & = & \delta_{xy}\sum_{t}\delta_{xx_{t}}(1+\gamma_{t})^{2}\sigma_{n}^{-2},\\
M_{xy,t} & = & 2\delta_{xy}\delta_{xx_{t}}(1+\gamma_{t})\sigma_{n}^{-2},\\
M_{xy,tt'} & = & 2\delta_{xy}\delta_{xx_{t}}\delta_{tt'}\sigma_{n}^{-2},\\
j_{x} & = & \sum_{t}(1+\gamma_{t})\,\delta_{xx_{t}}d_{t}\sigma_{n}^{-2},\\
j_{x,t} & = & \delta_{xx_{t}}d_{t}\sigma_{n}^{-2},\mbox{ and }j_{x,tt'}=0.
\end{eqnarray*}
}

The Gaussian signal, noise, and calibration covariances are assumed
to be known and to be described by power spectra in Fourier space.
In our concrete example, we use
\begin{eqnarray}
P_{s}(k) & = & \frac{a_{s}}{\left[1+(k/k_{s})^{2}\right]^{2}},\nonumber \\
P_{\gamma}(\omega) & = & \frac{a_{\gamma}}{\left[1+(\omega/\omega_{\gamma})^{2}\right]^{2}},\mbox{ and}\nonumber \\
P_{n}(\omega) & = & a_{n},\label{eq:spectra-example}
\end{eqnarray}
respectively. We express the amplitudes as $a_{s}=\sigma_{s}^{2}\lambda_{s}$,
$a_{\gamma}=\sigma_{\gamma}^{2}\tau_{\gamma}u$, and $a_{n}=\sigma_{n}^{2}\tau_{n}$
in terms of their respective variances $\sigma_{s}^{2}=\langle s_{x}^{2}\rangle_{(s)}$,
$\sigma_{\gamma}^{2}=\langle\gamma_{t}^{2}\rangle_{(\gamma)}$, and
$\sigma_{n}^{2}=\langle n_{t}^{2}\rangle_{(n)}$ and correlation lengths
$\lambda_{s}=4/k_{s}$, $\tau_{\gamma}=4/\omega_{\gamma}$, and $\tau_{n}=\tau$.
We choose $\sigma_{s}=1$, $\sigma_{\gamma}=0.75$, and $\sigma_{n}=0.2$
and correlation lengths $\lambda_{s}=0.3$ and $\tau_{\gamma}=1.5$.
This way, we have a unit variance signal, a 75\% calibration uncertainty
and 20\% white noise per measurement (in terms of typical signal strength).
The noise is white, the signal short-correlated (with about 3 correlation
regions within the signal domain $\Omega$) and the gain correlates
over a slightly larger region (a bit more than the size of the signal
domain $\Omega$). The gains are only slightly correlated between
subsequent passages over the same position ($\tau_{\gamma}=1.5$).

Any systematic difference in the data resulting from identical signal
positions should be due to gain variations. A decent \emph{selfcal}
scheme should be able to exploit this redundancy to estimate the gains
and therefore the signal. 

However, a global degeneracy of the data with respect to its variations
being caused by signal and gain variations can only partly be broken
by the three redundant scans over the signal domain. The data $d_{t}\approx(1+\gamma_{t})\, s_{x_{t}}$
only report a product of signal and response and one of those can
be traded of for the other. Therefore, a few external calibration
measurements are essential to break the degeneracy globally. 

To fix this degeneracy, we assume that four additional external calibration
measurements of the gain value have been performed at certain times
$t_{j}\in\{0,\,0.75,\,1.5,\,2.25\}$, with $d_{j}'=(1+\gamma_{t_{j}})\, c+n_{j}'$
by momentarily switching the observation to a strong calibration source
with a known strength of $c=4$. We assume that the noise during these
calibration measurements is as before, $n_{j}'\hookleftarrow\mathcal{G}(n_{j}',\,\sigma_{n}^{2})$.%
\footnote{For mental and notational convenience we ignore that during the external
calibration measurement usually no science signal data can be taken
by real instruments. However, this idealization is inessential and
has only a negligible impact on the results. %
}

The \emph{selfcal} equations become
\begin{eqnarray}
\gamma^{\star} & = & \Delta\, h,\mbox{ with}\label{eq:selcal}\\
\Delta_{tt'}^{-1} & = & \Gamma_{tt'}^{\text{-1}}+\delta_{tt'}\left(q_{t}+c^{2}\sum_{j}\delta_{tt_{j}}\right)\mbox{\ensuremath{\sigma_{n}^{-2}}},\nonumber \\
h_{t} & = & \left[d_{t}m_{x_{t}}-q_{t}+c^{2}\sum_{j}\delta_{tt_{j}}d_{j}'\right]\sigma_{n}^{-2},\mbox{ and}\nonumber \\
q_{t} & = & m_{x_{t}}^{2}+T\, D_{x_{t}x_{t}}.\nonumber 
\end{eqnarray}
Here we introduced the expected posterior variance of the signal realization
as constrained by the data, $q_{t}=\langle s_{x_{t}}^{2}\rangle_{(s|d,\gamma)}=m_{x_{t}}^{2}+D_{x_{t}x_{t}}$.
Furthermore we introduced $T$ as a parameter that switches between
classical \emph{selfcal} ($T=0$) and the new signal marginalized
\emph{selfcal} ($T=1$).

\subsection{Calibration comparison}

A simulated signal, gain, and resulting data realization using the
above specifications, as well as their reconstructions using different
information, assumptions, and approximations can be seen in Figs.~\ref{fig:Simulated-signal-and},
\ref{fig:Signal-reconstruction-and}, and \ref{fig:Signal-reconstruction-with_selfcal}.
These were generated using the generic signal inference framework
NIFT{\scriptsize Y}%
\footnote{To be found at \url{www.mpa-garching.mpg.de/ift/nifty}.%
} \citep{2013A&A...554A..26S}.

We quantify the signal and gain reconstructions in terms of their
average squared errors, $\varepsilon_{s}^{2}=(m-s)^{\dagger}(m-s)$
and $\varepsilon_{\gamma}^{2}=\frac{1}{u}(\gamma^{\star}-\gamma)^{\dagger}(\gamma^{\star}-\gamma)$,
respectively. Their expectation values, in case of known Gaussian
statistics, are given by
\begin{eqnarray}
\langle\varepsilon_{s}^{2}\rangle_{(d,s|\gamma)} & = & \int_{0}^{1}dx\, D_{xx}\mbox{ and}\nonumber \\
\langle\varepsilon_{\gamma}^{2}\rangle_{(d,\gamma|s)} & = & \frac{1}{3}\int_{0}^{3}dt\,\Delta_{tt}.
\end{eqnarray}

\begin{table}[t]
\caption{Reconstruction error of the different reconstruction for the example
shown in Figs.~\ref{fig:Simulated-signal-and}\protect\nobreakdash-\ref{fig:Signal-reconstruction-with_selfcal}.\label{tab:Reconstruction-error-of}}

\begin{tabular}{lr@{\extracolsep{0pt}.}lr@{\extracolsep{0pt}.}l}
\hline 
reconstruction method & \multicolumn{2}{c}{$\varepsilon_{s}$} & \multicolumn{2}{c}{$\varepsilon_{\gamma}$}\tabularnewline
\hline 
Wiener filter using known gains/signal & 0&037 & 0&056\tabularnewline
expected uncertainties of above & 0&040 & 0&063\tabularnewline
Gibbs sampling & 0&073 & 0&116\tabularnewline
expected uncertainty of above & 0&042 & 0&076\tabularnewline
no calibration, unit gains & 0&110 & 0&533\tabularnewline
only external calibration & 0&081%
\footnote{In this particular realization of signal, gain, and data, despite
the external calibration being relatively poor it has coincidentally
provided a better signal reconstruction than classical \emph{selfcal}.%
} & 0&246\tabularnewline
classical \emph{selfcal} & 0&089 & 0&192\tabularnewline
new \emph{selfcal} & 0&073 & 0&141\tabularnewline
\hline 
\end{tabular}
\end{table}

The best results are of course obtained when signal or calibration
are known. These Wiener filter solutions are optimal (dotted lines
in Fig.~\ref{fig:Signal-reconstruction-and}) and their uncertainty
estimates are reliable (gray regions in Fig.~\ref{fig:Signal-reconstruction-and}). 

The worst signal reconstruction is the one obtained while assuming
unit gains (thin gray lines in Fig.~\ref{fig:Signal-reconstruction-and}).
Using only the four external calibration measurements gives slightly
better results (dashed lines in Fig.~\ref{fig:Signal-reconstruction-and}).
The classical \emph{selfcal} provides more accurate calibration (dashed
lines in Fig.~\ref{fig:Signal-reconstruction-with_selfcal}), which
is further improved by the uncertainty corrections included in the
new \emph{selfcal} scheme (solid lines in Fig.~\ref{fig:Signal-reconstruction-with_selfcal}).
The best \emph{selfcal} solutions are provided by the Gibbs sampling.
Despite some numerical noise in the results, which can only be suppressed
by investing a large number of samples, these are\emph{ }optimal and
therefore provide a good benchmark for comparison. The \emph{new selfcal
}scheme obviously does not fully reach the accuracy of the Gibbs sampling.
Nevertheless, it is a significantly improvement over the \emph{classical
selfcal} as its solutions are visibly closer to the optimal Gibbs
sampling results.

These numbers and also the bottom panels of Fig.~\ref{fig:Signal-reconstruction-with_selfcal}
show further that the uncertainties in the calibration are systematically
larger than those of the signal. This is due to the fact that the
\emph{selfcal} has to rely on the signal being significantly non-zero,
which is not the case for many locations, whereas the signal reconstruction
is data driven for all positions except some rare points where the
gain $g=1+\gamma$ happens to vanish.

Since the signal uncertainty correction of the calibration removed
a systematic bias of the classical scheme, which had let to overestimated
gain solutions, the corresponding reconstructed signal shows more
variation as the one without this correction. This is visible by careful
inspection of the top left panels of Fig. \ref{fig:Signal-reconstruction-with_selfcal}.

\section{Conclusions\label{sec:Conclusions}}

We investigated the calibration problem of signal reconstruction from
data. Although we concentrated on simplified cases, approximating
all uncertainties in signal, calibration, and noise to be Gaussian
distributed, we believe that the gained qualitative insights are also
valid in many other circumstances. 

In case a perfect or sufficient external calibration measurement is
missing, the signal to be measured has also to serve as a calibrator.
This is usually done by \emph{selfcal} schemes, which reconstruct
the signal assuming some calibration, calibrate on the reconstructed
signal, and iterate this until convergence or other termination criteria
are met. We have shown that such \emph{selfcal} schemes arise naturally
from trying to maximize the joint posterior of signal and calibration.
We therefore demonstrated that any fix point of such \emph{selfcal}
iterations must be a maximum of this posterior. There is, however,
no guarantee that the obtained maximum is a global one. 

The joint MAP estimator is not necessarily optimal in the sense of
an minimal expected square error. Due to the interwoven coupling between
signal and calibration in the data this maximum is indeed not optimal.
In order to obtain improved signal and calibration schemes, we worked
out the calibration marginalized signal posterior and the signal marginalized
calibration posterior and the resulting maximum a posteriori estimators.
Both contain corrections terms taking into account the remaining uncertainties
of calibration and signal, respectively. 

For the canonical situation that the signal is a quantity that varies
around zero, whereas the signal response has a known non-zero part,
we argue that the calibration corrections due to signal uncertainties
are more essential than the signal reconstruction corrections due
to calibration uncertainties. The reason is that in this case, the
information source of the data on the unknown signal contains a calibration
independent term, whereas the information source for the calibration
requires signal information. This is reflected by the observation
that the calibration uncertainty corrections for the signal as given
by Eq.~\eqref{eq:calib marginalized signal} contain pairs of mutually
nearly canceling terms. 

In contrast to this, the calibration correction for signal uncertainties
as given by Eq. \eqref{eq:asdjfkla} is of a systematic nature. It
reduces, on average and for positive known part of the response, the
values of the inferred calibration solution. This leads to a more
pronounced and thereby more accurate signal reconstruction as more
of the data variance can be assigned to the signal. We have illustrated
this with a simplistic numerical example. 

The proposed improvement of \emph{selfcal} schemes should not be regarded
as the ultimate theory of calibration. A number of approximations
have been incorporated in order to limit the computational complexity.
In particular the mutual dependence of signal and calibration uncertainties
are not fully taken into account and only the dominant influence of
the uncertainties on the posterior means of signal and calibration
were calculated. A comparison with a numerically expensive, but asymptotically
exact Gibbs sampling scheme shows that the corrections are indeed
a good step in the right direction. However, they also show that there
is still space for further improvements. 

Thus, we believe that these corrections can help to refine the contemporary
art of calibration and thereby improve measurement results in many
areas of science and technology.

\subsection*{Acknowledgements}

We gratefully thank Vanessa Böhm, Sebastian Dorn, and Niels Opermann,
for discussions, feedback, and comments. The calculations were performed
using the generic signal inference framework NIFT{\scriptsize Y} to
be found at \url{www.mpa-garching.mpg.de/ift/nifty} \citep{2013A&A...554A..26S}.

\bibliographystyle{apsrev4-1}
\bibliography{../../Text/Bib/ift,calib}

\end{document}